\documentclass[aps,prd,preprint,showpacs,nofootinbib]{revtex4-1}

\usepackage{amsfonts,amssymb,amsmath,dsfont,graphicx}

\begin{document}


\title{Status of non-standard neutrino interactions}


\author{Tommy Ohlsson}
\email[]{tohlsson@kth.se}
\affiliation{Department of Theoretical Physics, School of Engineering Sciences, KTH Royal Institute of Technology -- AlbaNova University Center, Roslagstullsbacken 21, 106 91 Stockholm, Sweden}


\date{\today}

\begin{abstract}
The phenomenon of neutrino oscillations has been established as the leading mechanism behind neutrino flavor transitions, providing solid experimental evidence that neutrinos are massive and lepton flavors are mixed.
Here we review sub-leading effects in neutrino flavor transitions known as non-standard neutrino interactions, which is currently the most explored description for effects beyond the standard paradigm of neutrino oscillations.
In particular, we report on the phenomeno\-logy of non-standard neutrino interactions and their experimental and phenomenological bounds as well as an outlook for future sensitivity and discovery reach.
\end{abstract}


\maketitle

\tableofcontents

\section{Introduction}
\label{sec:intro}

Since the results of the Super-Kamiokande experiment in Japan in 1998 \cite{Fukuda:1998mi}, the phenomenon of neutrino oscillations has been established as the leading mechanism behind neutrino flavor transitions. This result was followed by a first boom of results from several international collaborations (e.g.~SNO, KamLAND, K2K, MINOS, and MiniBooNE) on the various neutrino parameters. Certainly, these solid results have pinned down the values on the different parameters to an incredible precision given that neutrinos are very elusive particles and the corresponding experiments are extraordinarily complex (see \cite{Ohlsson:2012}). Nevertheless, it is a fact that the present Standard Model (SM) of particle physics is not the whole story and needs to be revised in order to accommodate massive and mixed neutrinos, which leads to physics beyond the SM. With the upcoming results from the running or future neutrino experiments (e.g.~Daya Bay, Double Chooz, ICARUS, IceCube, KATRIN, NO$\nu$A, OPERA, RENO, T2K)\footnote{Note that no experiments on neutrinoless double beta decay have been included in the list of examples. For a recent review on the physics of neutrinoless double beta decay, see, e.g., \cite{Rodejohann:2011mu}, and especially, section~V for non-standard interactions in connection with neutrinoless double beta decay.}, there will be a second boom of results, and we will hopefully be able to determine the missing neutrino parameters such as the sign of the large mass-squared difference for neutrinos (important for the neutrino mass hierarchy), the leptonic CP-violating phase (important for the matter-antimatter asymmetry in the Universe), and the absolute neutrino mass scale (using the KATRIN experiment), but also the next-to-leading order effects in neutrino flavor transitions.

In future neutrino experiments (and in particular for a neutrino factory, $\beta$-beams, or superbeams), `new physics' beyond the SM may appear in the form of unknown couplings involving neutrinos, which are usually referred to as non-standard neutrino interactions (NSIs). Compared with standard neutrino oscillations, NSIs could contribute to the oscillation probabilities and neutrino event rates as sub-leading effects, and may bring in very distinctive phenomena. Running and future neutrino experiments will provide us with more precision measurements on neutrino flavor transitions, and therefore, the window of searching for NSIs is open. In principle, NSIs could exist in the neutrino production, propagation, and detection processes, and the search for NSIs is complementary to the direct search for new physics conducted at the LHC.
The main motivation to study NSIs is that if they exist we ought to know their effects on physics. Models of physics that predict NSIs include, for example, various seesaw models, R-parity violating supersymmetric models, left-right symmetric models, GUTs, and extra dimensions, i.e.~basically all modern models for physics beyond the SM could give rise to NSIs. For some specific models, see section~\ref{sec:NSImodels} and references therein.

The concept of NSIs has been introduced in order to accommodate for sub-leading effects in neutrino flavor transitions. Previously, alternative scenarios for neutrino flavor transitions such as neutrino decoherence, neutrino decay and NSIs have been studied (see below for references), but now, such alternatives are only allowed to provide sub-leading effects to neutrino oscillations. In the literature, there exist several theoretical and phenomenological studies of NSIs for atmospheric, accelerator, reactor, solar and supernova neutrinos (see especially the references given in section~\ref{sec:NSIpheno}). In addition, some experimental collaborations have obtained bounds on NSIs (see \cite{Mitsuka:2011ty,Coelho:2012}). The different types of experiments that are relevant to NSIs include, e.g., neutrino oscillations experiments, experiments on lepton flavor violating processes, experiments on neutrino cross-sections and data from experiments at accelerators (such as the LEP collider, the Tevatron and the LHC).

This review is organized as follows. In section~\ref{sec:NSIs}, we introduce the concept of neutrino flavor transitions with NSIs. First, we present standard neutrino oscillations, and then, we consider other scenarios for neutrino flavor transitions including NSIs. At the end of the section, we discuss so-called NSI Hamiltonian effects of neutrino oscillations. Then, in section~\ref{sec:NSIs3}, we describe NSIs with three neutrino flavors, since there are at least three flavors in Nature. Especially, we consider production, propagation, and detection NSIs including the so-called zero-distance effect. In addition, we present mappings for NSIs and approximate formulas for two neutrino flavors that can be useful in some settings. Next, in section~\ref{sec:NSImodels}, we study different theoretical models for NSIs including, e.g., a seesaw model. In section~\ref{sec:NSIpheno}, we investigate the phenomenology of NSIs for different types of neutrinos such as atmospheric, accelerator, reactor, solar and supernova neutrinos, whereas in section~\ref{sec:NSIexp}, we review phenomenological bounds on NSIs. In addition, in section~\ref{sec:NSIdiscovery}, we give an outlook and examine experimental sensitivities and the future discovery reach of NSIs. Finally, in section~\ref{sec:S&C}, we present a summary and state our conclusions.

\section{Neutrino flavor transitions with NSIs}
\label{sec:NSIs}

In this section, we present the basic ingredients for neutrino oscillations (based on the two facts that neutrinos are massive and lepton flavors are mixed), which is the leptonic mixing matrix, a Schr{\"o}dinger-like equation for the evolution of the neutrinos, and the values of the fundamental neutrino oscillation parameters, i.e.~the neutrino mass-squared differences and the leptonic mixing parameters. Then, we discuss some historic alternative scenarios for neutrino flavor transitions. Finally, we study NSI Hamiltonian effects of neutrino oscillations.

\subsection{Neutrino oscillations}

Indeed, there are now strong evidences that neutrinos are massive and lepton flavors are mixed. In the SM, neutrinos are massless particles, and therefore, the SM must be extended by adding neutrino masses. The lepton flavor mixing is usually defined through the leptonic mixing matrix $U$ that can be written as \cite{Pontecorvo:1957cp,Pontecorvo:1957qd,Maki:1962mu,Pontecorvo:1967fh}
\begin{equation}
\left( \begin{matrix} \nu_e \\ \nu_\mu \\ \nu_\tau \end{matrix} \right) = U \left( \begin{matrix} \nu_1 \\ \nu_2 \\ \nu_3 \end{matrix} \right) = \left( \begin{matrix} U_{e1} & U_{e2} & U_{e3} \\ U_{\mu 1} & U_{\mu 2} & U_{\mu 3} \\ U_{\tau 1} & U_{\tau 2} & U_{\tau 3} \end{matrix} \right) \left( \begin{matrix} \nu_1 \\ \nu_2 \\ \nu_3 \end{matrix} \right) \,,
\end{equation}
which relates the weak interaction eigenstates and the mass eigenstates through the leptonic mixing parameters $\theta_{12}$, $\theta_{13}$, $\theta_{23}$, $\delta$ (the Dirac CP-violating phase), as well as $\rho$ and $\sigma$ (the Majorana CP-violating phases). In the so-called standard parameterization, $U$ is given by \cite{Beringer:1900zz}
\begin{equation}
U = \left( \begin{matrix} 1 & 0 & 0 \\ 0 & c_{23} & s_{23} \\ 0 & -s_{23} & c_{23} \end{matrix} \right) \left( \begin{matrix} c_{13} & 0 & s_{13} {\rm e}^{-{\rm i} \delta} \\ 0 & 1 & 0 \\ -s_{13} {\rm e}^{{\rm i} \delta} & 0 & c_{13} \end{matrix} \right) \left( \begin{matrix} c_{12} & s_{12} & 0 \\ -s_{12} & c_{12} & 0 \\ 0 & 0 & 1 \end{matrix} \right) \left( \begin{matrix} {\rm e}^{{\rm i} \rho} & 0 & 0 \\ 0 & {\rm e}^{{\rm i} \sigma} & 0 \\ 0 & 0 & 1 \end{matrix} \right) \,,
\end{equation}
where $c_{ij} \equiv \cos (\theta_{ij})$ and $s_{ij} \equiv \sin (\theta_{ij})$.

The time evolution of the neutrino vector of state $\nu = \left( \begin{matrix} \nu_e & \nu_\mu
& \nu_\tau \end{matrix} \right)^T$ describing neutrino oscillations is given by a Schr{\"o}dinger-like equation with a Hamiltonian $H$, namely, (see, e.g., \cite{Bilenky:1987ty} for a detailed review)
\begin{equation}
{\rm i} \frac{{\rm d}\nu}{{\rm d} t} = \frac{1}{2E} \left[ M M^\dagger
+ {\rm diag}(A,0,0) \right] \nu \equiv H \nu \,,
\label{eq:sch}
\end{equation}
where $E$ is the neutrino energy, $M = U \, {\rm diag}(m_1,m_2,m_3) \, U^T$ is the neutrino mass matrix, and $A = 2 \sqrt{2} E G_F N_e$ is the effective matter potential induced by ordinary charged-current weak interactions with electrons \cite{Wolfenstein:1977ue,Mikheev:1986gs}. Here, $m_1$, $m_2$, and $m_3$ are the definite masses of the neutrino mass eigenstates, $G_F = (1.1663787 \pm 0.0000006) \times 10^{-5} \, {\rm GeV}^{-2}$ is the Fermi coupling constant \cite{Beringer:1900zz}, and $N_e$ is the electron density of matter along the neutrino trajectory. Quantum mechanically, the transition probability amplitudes are given as overlaps of different neutrino states, and finally, neutrino oscillation probabilities are defined as squared absolute values of the transition probability amplitudes. Thus, flavor transitions occur during the evolution of neutrinos. For example, in a two-flavor illustration (in vacuum) with electron and muon neutrinos, a neutrino state can be in a pure electron neutrino state at one time, whereas it can be in a pure muon neutrino state at another time. In this case, the well-known two-flavor neutrino oscillation probability formulae are given by (see, e.g., \cite{Bilenky:1987ty})
\begin{eqnarray}
P(\nu_e \to \nu_\mu;L) &=& \sin^2 (2\theta) \sin^2 \left( \frac{\Delta m^2 L}{4E} \right) \,, \label{eq:2nu}\\
P(\nu_e \to \nu_e) &=& 1-P(\nu_e \to \nu_\mu) = 1-P(\nu_\mu \to \nu_e) = P(\nu_\mu \to \nu_\mu)\,,
\end{eqnarray}
where $L$ is the (propagation) path length of the neutrinos, $\theta$ is the two-flavor mixing angle (corresponding to the amplitude of the oscillations) and $\Delta m^2$ is the mass-squared difference (corresponding to the frequency of the oscillations) between the masses of the two neutrino mass eigenstates. In addition, in the case of three neutrino flavors in vacuum, we have the more cumbersome formula for the neutrino transition probability
\begin{eqnarray}
P(\nu_\alpha \to \nu_\beta;L) & = & \delta_{\alpha\beta} - 4 \sum_{i>j} {\rm Re} ( U^*_{\alpha i} U_{\beta i} U_{\alpha j} U^*_{\beta j} )\sin^2\left(\frac{\Delta m^2_{ij}L}{4E}\right) \nonumber\\
& + & 2 \sum_{i>j}{\rm Im} ( U^*_{\alpha i} U_{\beta i} U_{\alpha j} U^*_{\beta j} ) \sin\left(\frac{ \Delta m^2_{ij} L}{2 E}\right) \,, \label{eq:3nu}
\end{eqnarray}
where $\alpha,\beta = e, \mu, \tau$. In fact, it even turns out that equation~(\ref{eq:3nu}) holds for arbitrary neutrino flavors.

Using global fits to data from neutrino oscillation experiments, the values given in table~\ref{tab:dataparam} have been obtained for the fundamental neutrino oscillation parameters \cite{GonzalezGarcia:2012sz}.
\begin{table}
\begin{tabular}{lccc}
\hline
Parameter & Best-fit value & $3\sigma$ range\\
\hline
$\Delta m_{21}^2$ [$10^{-5} \, {\rm eV}^2$] & $7.50 \pm 0.185$ & $7.00 \div 8.09$ \\
$|\Delta m_{31}^2|$ [$10^{-3} \, {\rm eV}^2$] & $2.47^{+0.069}_{-0.067}$ & $2.27 \div 2.69$ \\
$\sin^2 (\theta_{12})$ & $0.30 \pm 0.013$ & $0.27 \div 0.34$ \\
$\sin^2 (\theta_{13})$ & $0.023 \pm 0.0023$ & $0.016 \div 0.030$ \\ 
$\sin^2 (\theta_{23})$ & $0.41^{+0.037}_{-0.025}$ & $0.34 \div 0.67$ \\
\hline
\end{tabular}
\caption{Present values of the fundamental neutrino oscillation parameters obtained in a global fit analysis using all available neutrino oscillation data \cite{GonzalezGarcia:2012sz}. See also \cite{Tortola:2012te,Fogli:2012ua} for two other analyses.}
\label{tab:dataparam}
\end{table}
Note that these values have been found without taking sub-leading effects such as NSIs into account. Open questions that still exist about neutrinos are: Are neutrinos Dirac or Majorana particles? What is the absolute neutrino mass scale? What is the sign of the large mass-squared difference $\Delta m_{31}^2$?\footnote{Note that the sign of the large mass-squared difference will determine the character of the neutrino mass spectrum, i.e.~if the spectrum follows normal or inverted neutrino mass hierarchy.} Is there leptonic CP violation? Do sterile neutrinos exist? However, recently, one has also been concerned with the following two questions in the literature: Are there NSIs? Is there non-unitarity in leptonic mixing? The intention of this review is to bring some insight into these last two questions (with emphasis on the first question).

Since 1998, the Super-Kamiokande, SNO and KamLAND experiments have provided strong evidence for neutrino flavor transitions and that the theory of neutrino oscillations is the leading description \cite{Fukuda:1998mi,Ahmad:2001an,Eguchi:2002dm}. In various neutrino oscillation experiments, precision measurements for some of the neutrino parameters, i.e.~$\Delta m_{21}^2$, $|\Delta m_{31}^2|$, $\theta_{12}$, $\theta_{13}$ and $\theta_{23}$, have been obtained, whereas other parameters are still completely unknown such as ${\rm sign}(\Delta m_{31}^2)$ and $\delta$, as well as the Majorana CP-violating phases and the absolute neutrino mass scale. Running and future neutrino oscillation experiments might have sensitivies to measure ${\rm sign}(\Delta m_{31}^2)$ and possibly $\delta$, while neutrinoless double beta decay experiments could determine if neutrinos are Dirac or Majorana particles (as well as the Majorana CP-violating phases) and the KATRIN experiment will probe the absolute neutrino mass scale using $\beta$-decay. New physics, such as NSIs, might be present and complicate the experiments that want to answer the fundamental questions about neutrinos. Thus, we should investigate NSIs in order to obtain knowledge on their possible effects.

\subsection{Other scenarios for neutrino flavor transitions}
\label{sub:other_scenarios}

Other mechanisms could be responsible for flavor transitions on a sub-leading level (see, e.g., \cite{Blennow:2005yk}). Therefore, we will phenomenologically study new physics effects due to NSIs. In the past, descriptions for transitions of neutrinos based on neutrino decoherence and neutrino decay have been extensively investigated in the literature \cite{Grossman:1998jq,Lisi:2000zt,Adler:2000vfa,Gago:2000qc,Gago:2000nv,Ohlsson:2000mj,Gago:2002na,Fogli:2003th,Barenboim:2004wu,Morgan:2004vv,Anchordoqui:2005gj,Barenboim:2006xt,Fogli:2007tx,Alexandre:2007na,Ribeiro:2007jq,Mavromatos:2007hv,Farzan:2008zv,Bahcall:1972my,Barger:1981vd,Valle:1983ua,Barger:1998xk,Choubey:1999ir,Barger:1999bg,Fogli:1999qt,Pakvasa:1999ta,Choubey:2000an,Bandyopadhyay:2001ct,Lindner:2001fx,Lindner:2001th,Joshipura:2002fb,Beacom:2002cb,Bandyopadhyay:2002qg,Beacom:2002vi,Indumathi:2002qb,Ando:2003ie,Fogli:2004gy,Ando:2004qe,PalomaresRuiz:2005vf,Meloni:2006gv,GonzalezGarcia:2008ru,Maltoni:2008jr,Mehta:2011qb,Baerwald:2012kc}. However, now, such descriptions are ruled out by available neutrino data as the leading-order mechanism behind neutrino flavor transitions \cite{Ashie:2004mr,Araki:2004mb,Fogli:2007tx,Adamson:2008zt,Adamson:2011ig}, but these descriptions could still provide sub-leading effects. In what follows, we will not consider neutrino decoherence and neutrino decay, but instead focus on NSIs, which are interactions between neutrinos and matter fermions (i.e.~$u$, $d$ and $e$) that additionally affect neutrino oscillations, as a sub-leading mechanism for neutrino flavor transitions. 

\subsection{NSI Hamitonian effects of neutrino oscillations}
\label{sub:NSI_H_eff}

In general, NSIs can be considered to be effective additional contributions to the standard vacuum Hamiltonian $H_0$ that describes the neutrino evolution (see, e.g., \cite{Blennow:2005qj} for details)\footnote{It should be noted that the idea of NSIs was first presented in the seminal work by Wolfenstein \cite{Wolfenstein:1977ue}. Other important works on NSIs can be found in \cite{Mikheev:1986gs,Valle:1987gv,Guzzo:1991hi,Roulet:1991sm,Brooijmans:1998py,GonzalezGarcia:1998hj,Bergmann:2000gp,Guzzo:2000kx,Guzzo:2001mi}.}. Thus, any Hermitian non-standard Hamiltonian effect $H'$ will alter the original Hamiltonian into an effective Hamiltonian:
\begin{equation}
H_{\rm eff} = H_0 + H' \,.
\end{equation}
For example, neutrino oscillations in matter with $1 < n \leq 3$ flavors, which is the canonical example of NSIs, are described by
\begin{eqnarray}
H' = H_{\rm matter} &=& \frac{1}{2E} {\rm diag}(A,0,\ldots,0) - \frac{1}{\sqrt{2}} G_F N_n \mathds{1}_n \nonumber\\
&=& \sqrt{2} G_F {\rm diag}(N_e - \tfrac{1}{2} N_n, - \tfrac{1}{2} N_n, \ldots, - \tfrac{1}{2} N_n)\,,
\end{eqnarray}
where the quantity $A$ was defined in connection to equation~(\ref{eq:sch}), $N_n$ is the nucleon number density and $\mathds{1}_n$ is the $n \times n$ unit matrix. Note that the opposite signs of the charged-current weak interaction contribution (proportional to $N_e$) and the neutral-current weak interaction contribution (proportional to $N_n$). In the case of neutrino oscillations in matter, the effective Hamiltonian $H_{\rm eff} = H_0 + H_{\rm matter}$ is basically the same Hamiltonian as the one defined in equation~(\ref{eq:sch}), since the neutral-current weak interaction contribution appears in all diagonal elements of the second term in $H_{\rm matter}$, which means that this term will only affect the phase of the time evolution, and therefore has no effect on neutrino oscillations.
Just as the presence of matter affects the effective neutrino parameters, the effective neutrino para\-meters will be affected by any non-standard Hamiltonian effect. For example, in the case of so-called matter NSIs---a generalization of neutrino oscillations in matter, the corresponding effective Hamiltonian will be presented and discussed in section~\ref{sub:mNSIs}.

In general, the non-standard Hamiltonian effects can alter both the oscillation frequency and the oscillation amplitude and they can be classified as `flavor effects' or `mass effects' \cite{Blennow:2005qj}. A non-standard Hamiltonian effect can be defined in either flavor or mass basis, and be parametrized by the so-called generators that span the effective Hamiltonian $H_{\rm eff}$ in the basis under consideration. If $n = 2$, the generators are the three Pauli matrices, whereas if $n = 3$, the generators are instead the eight Gell-Mann matrices. For example, NSIs and flavor-changing neutral currents \cite{Wolfenstein:1977ue,Krastev:1997cp} are normally defined in flavor basis, whereas the concept of mass-varying neutrinos \cite{Gu:2003er,Fardon:2003eh} is defined in mass basis. In principle, there is no mathematical difference between flavor and mass effects if one allows for the most general form in each basis. However, one can define a non-standard Hamiltonian effect as a `pure' flavor or mass effect if the corresponding Hamiltonian $H'$ can be written as $H' = c \rho_i$ or $H' = c \tau_i$ ($i$ fixed), where $c$ is a real number and $\rho_i$'s and $\tau_i$'s are the generators in flavor basis and mass bases, respectively. Thus, pure effects are restricted to be of very specific types, where the actual forms are very simple in either flavor or mass basis, and correspond to pure flavor/mass conserving/violating effects, i.e.~effects that affect particular flavor or mass eigenstates.

Furthermore, non-standard Hamiltonian effects (such as NSIs) will lead to resonance conditions \cite{Blennow:2005qj}, which are modified versions of the famous {\it Mikheyev--Smirnov--Wolfenstein (MSW) effect} \cite{Wolfenstein:1977ue,Mikheev:1986gs,Mikheev:1986wj}. See, e.g., section~\ref{sub:approx2fl}.

\section{NSIs with three neutrino flavors}
\label{sec:NSIs3}

The phenomenological consequences of NSIs have been investigated in great detail in the literature. The widely studied operators responsible for NSIs can be written as \cite{Wolfenstein:1977ue,Grossman:1995wx,Berezhiani:2001rs,Davidson:2003ha}
\begin{equation}
{\cal L}_{\rm NSI} = -2 \sqrt{2} G_F \varepsilon_{\alpha\beta}^{ff'C} \left( \overline{\nu_\alpha} \gamma^\mu P_L \nu_\beta \right) \left( \overline{f} \gamma_\mu P_C f' \right) \,,
\label{eq:NSIops}
\end{equation}
where $\varepsilon_{\alpha\beta}^{ff'C}$ are NSI parameters, $\alpha,\beta = e,\mu,\tau$, $f,f' = e,u,d$ and $C = L,R$. If $f \neq f'$, the NSIs are charged-current like, whereas if $f = f'$, the NSIs are neutral-current like and the NSI parameters are defined as $\varepsilon_{\alpha\beta}^{fC} \equiv \varepsilon_{\alpha\beta}^{ffC}$. Note that the operators~(\ref{eq:NSIops}) are non-renormalizable and they are also not gauge invariant. Thus, using the NSI operators in equation~(\ref{eq:NSIops}), which lead to a so-called dimension-6 operator after heavy degrees of freedom are integrated out, and the well-known relation $G_F/\sqrt{2} \simeq g_W^2/(8 m_W^2)$,\footnote{The quantity $g_W$ is the coupling constant of the weak interaction.} we find that the effective NSI parameters are (see, e.g., \cite{GonzalezGarcia:2001mp,Kopp:2007ne,Minakata:2008gv} for discussions)
\begin{equation}
\varepsilon \propto \frac{m_W^2}{m_X^2} \,,
\label{eq:mW2mX2}
\end{equation}
where $m_W = (80.385 \pm 0.015) \, {\rm GeV} \sim 0.1 \, {\rm TeV}$ is the W boson mass and $m_X$ is the mass scale at which the NSIs are generated \cite{Beringer:1900zz}. Note that the characteristic proportionality relation~(\ref{eq:mW2mX2}) is at least valid for energies below the new physics scale $m_X$, where the NSI operators are effective. If the new physics scale, i.e.~the NSI scale, is of the order of 1(10)~TeV, then one obtains effective NSI parameters of the order of $\varepsilon_{\alpha\beta} \sim 10^{-2} (10^{-4})$.

In principle, NSIs can affect both (i) production and detection processes and (ii) propagation in matter and (iii) one can have combinations of both effects. In the following, we will first study production and detection NSIs, including the so-called zero-distance effect, and then matter NSIs. In addition, we will present mappings with NSIs and discuss approximate formulae for two neutrino flavors.

\subsection{Production and detection NSIs and the zero-distance effect}
\label{sub:pdNSIs}

In general, production and detection processes, which are based on charged-current interaction processes, can be affected by charged-current like NSIs. For a realistic neutrino oscillation experiment, the neutrino states produced in a source and observed at a detector can be written as superpositions of pure orthonormal flavor eigenstates \cite{Grossman:1995wx,GonzalezGarcia:2001mp,Bilenky:1992wv,Meloni:2009cg}:
\begin{eqnarray}\label{eq:s}
|\nu^{\rm s}_\alpha \rangle & = &  |\nu_\alpha \rangle +
\sum_{\beta=e,\mu,\tau} \varepsilon^{\rm s}_{\alpha\beta}
|\nu_\beta\rangle  = (1 + \varepsilon^{\rm s}) U  |\nu_m \rangle \,, \\
\langle \nu^{\rm d}_\beta| & = &\langle
 \nu_\beta | + \sum_{\alpha=e,\mu,\tau}
\varepsilon^{\rm d}_{\alpha \beta} \langle  \nu_\alpha  | =  \langle \nu_m
|  U^\dagger [1 + (\varepsilon^{\rm d})^\dagger] \,, \label{eq:d}
\end{eqnarray}
where the superscripts `s' and `d' denote the source and the detector, respectively, and $|\nu_m\rangle$ is a neutrino mass eigenstate. In addition, the production and detection NSI parameters, i.e.~$\varepsilon^{\rm s}_{\alpha\beta}$ and $\varepsilon^{\rm d}_{\alpha\beta}$, are defined through NSI parameters $\varepsilon_{\alpha\beta}^{ff'C}$, where $f \neq f'$. Note that the states $|\nu_\alpha^{\rm s} \rangle$ and $\langle \nu_\beta^{\rm d}|$ are not orthonormal states due to the NSIs and that the matrices $\varepsilon^{\rm s}$ and $\varepsilon^{\rm d}$ are not necessarily the same matrix, since different physical processes take place at the source and the detector, which means that these matrices are arbitrary and non-unitary in general. If the production and detection processes are exactly the same process with the same participating fermions (e.g.~$\beta$-decay and inverse $\beta$-decay), then the same matrix enters as $\varepsilon^{\rm s} = \left(\varepsilon^{\rm d}\right)^\dagger$,  or on the form of matrix elements, $\varepsilon_{\alpha\beta}^{\rm s} = \varepsilon_{\alpha\beta}^{\rm d} = (\varepsilon_{\beta\alpha}^{\rm s})^* = (\varepsilon_{\beta\alpha}^{\rm d})^*$ \cite{Kopp:2007ne}. For example, in the case of so-called non-unitarity effects (which can be considered as a type of NSIs, see, e.g., \cite{Xing:2008hx}) in the minimal unitarity violation model \cite{Antusch:2006vwa,FernandezMartinez:2007ms,Goswami:2008mi,Xing:2008fg,Luo:2008vp,Altarelli:2008yr}, it holds that $\varepsilon^{\rm s} = \left(\varepsilon^{\rm d}\right)^\dagger$. Thus, it is important to keep in mind that these matrices are experiment- and process-dependent quantities.

In the case of production and detection NSIs, the neutrino transition probabilities are given by (see equation~(\ref{eq:3nu}) for the case without production and detection NSIs) \cite{Ohlsson:2008gx,Meloni:2009cg}
\begin{eqnarray}
P(\nu_\alpha^{\rm s} \to \nu_\beta^{\rm d};L) & = & \left| \sum_{\gamma,\delta,i} \left( 1 + \varepsilon^{\rm d} \right)_{\gamma\beta} \left(1 + \varepsilon^{\rm s} \right)_{\alpha\delta} U_{\delta i} U^*_{\gamma i} \, {\rm e}^{-{\rm i} \frac{m_i^2 L}{2E}} \right|^2 \nonumber\\
& = &  \sum_{i,j} {\cal J}^i_{\alpha\beta} {\cal J}^{j*}_{\alpha\beta} - 4 \sum_{i>j} {\rm Re} ({\cal J}^i_{\alpha\beta} {\cal J}^{j*}_{\alpha\beta} )\sin^2\left(\frac{\Delta m^2_{ij}L}{4E}\right) \nonumber\\
& + & 2 \sum_{i>j}{\rm Im} ( {\cal J}^i_{\alpha\beta} {\cal J}^{j*}_{\alpha\beta} ) \sin\left(\frac{ \Delta m^2_{ij} L}{2 E}\right) \,, \label{eq:Pnusnud}
\end{eqnarray}
where
\begin{equation}
{\cal J}^i_{\alpha\beta}  = {U^*_{\alpha i} U_{\beta i} + \sum_\gamma \varepsilon^{\rm s}_{\alpha \gamma} U^*_{\gamma i} U_{\beta i}+ \sum_\gamma \varepsilon^{\rm d}_{\gamma \beta} U^*_{\alpha i} U_{\gamma i}  + \sum_{\gamma,\delta} \varepsilon^{\rm s}_{\alpha\gamma} \varepsilon^{\rm d}_{\delta \beta} U^*_{\gamma i} U_{\delta i} } \,.
\end{equation}
In fact, an important feature of equation~(\ref{eq:Pnusnud}) is that the first term, i.e.~$\sum_{i,j} {\cal J}^i_{\alpha\beta} {\cal J}^{j*}_{\alpha\beta}$, is generally different from zero or one. Especially, evaluating equation~(\ref{eq:Pnusnud}) at $L = 0$, we obtain
\begin{equation}
P(\nu_\alpha^{\rm s} \to \nu_\beta^{\rm d};L = 0) =  \sum_{i,j} {\cal J}^i_{\alpha\beta} {\cal J}^{j*}_{\alpha\beta} \,,
\end{equation}
which means that a neutrino flavor transition would already happen at the source before the oscillation process has taken place. This is known as the {\it zero-distance effect} \cite{Langacker:1988up}. It could be measured with a near detector close to the source. In the case that $\varepsilon^{\rm s} = \varepsilon^{\rm d} = 0$, i.e.~without production and detection NSIs, the first term reduces to
\begin{equation}
\sum_{i,j} {\cal J}^i_{\alpha\beta} {\cal J}^{j*}_{\alpha\beta} = \sum_{i,j} U^*_{\alpha i} U_{\beta i} U_{\alpha j} U^*_{\beta j} = \delta_{\alpha\beta} \,,
\end{equation}
which is the first term in equation~(\ref{eq:3nu}). Note that equation~(\ref{eq:Pnusnud}) is also usable to describe neutrino oscillations with a non-unitary mixing matrix, e.g.~in the minimal unitarity violation model \cite{Antusch:2006vwa}.

\subsection{Matter NSIs}
\label{sub:mNSIs}

\begin{figure}
  \includegraphics[width=0.95\textwidth]{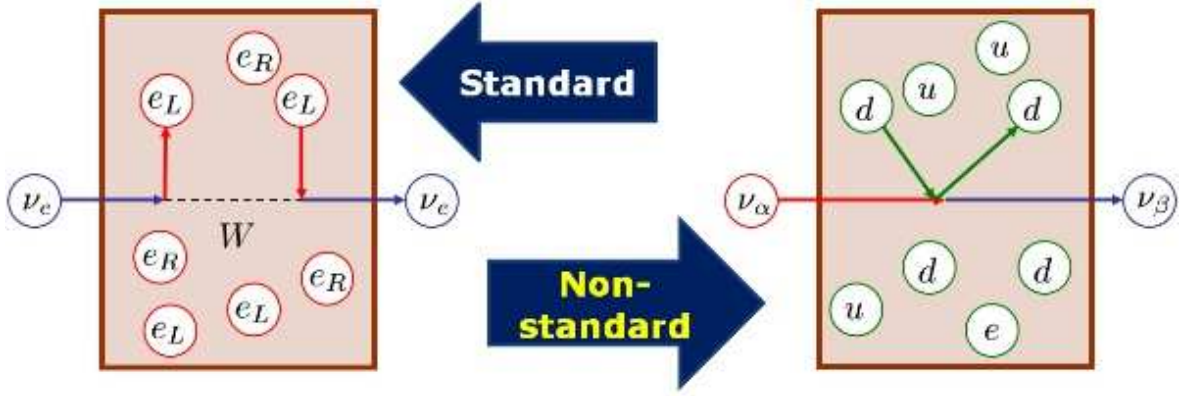}
  \vspace{-5mm}
  \caption{Schematic pictures of standard matter effects (left picture) and matter non-standard neutrino interactions (right picture).}
  \label{fig:NSIschematic}
\end{figure}

In order to describe neutrino propagation in matter with NSIs (assuming no effect of production and detection NSIs, which were discussed in section~\ref{sub:pdNSIs}), the simple effective matter potential in equation~(\ref{eq:sch}) needs to be extended. Similar to standard matter effects, NSIs can affect the neutrino propagation by coherent forward scattering in Earth matter. The Earth matter effects are more or less involved depending on the specific terrestrial neutrino oscillation experiment. In other words, the Hamiltonian in equation~(\ref{eq:sch}) is replaced by an effective Hamiltonian, which governs the propagation of neutrino flavor states in matter with NSIs, namely \cite{Wolfenstein:1977ue,Valle:1987gv,Guzzo:1991hi,Roulet:1991sm}
\begin{equation}
\hat{H} = \frac{1}{2E} \left[ U {\rm diag}(m_1^2,m_2^2,m_3^2) U^\dagger + {\rm diag}(A,0,0) + A\varepsilon^m \right] \,, \label{eq:Heff}
\end{equation}
where the matrix $\varepsilon^m$ contains the (effective) matter NSI parameters $\varepsilon_{\alpha\beta}$ ($\alpha,\beta = e,\mu,\tau$), which are defined as
\begin{equation}
\varepsilon_{\alpha\beta} \equiv \sum_{f,C} \varepsilon_{\alpha\beta}^{fC} \frac{N_f}{N_e}
\label{eq:matterNSIs}
\end{equation}
with the parameters $\varepsilon_{\alpha\beta}^{fC}$ being entries of the Hermitian matrix $\varepsilon^{fC}$ and giving the strengths of the NSIs and the quantity $N_f$ being the number density of a fermion of type $f$. Unlike $\varepsilon^{\rm s}$ and $\varepsilon^{\rm d}$, $\varepsilon^m = (\varepsilon_{\alpha\beta})$ is a Hermitian matrix describing NSIs in matter, where the superscript `$m$' is used to distinguish matter NSIs from production and detection NSIs. Thus, for three neutrino flavors, we obtain
\begin{equation}
{\rm i} \frac{{\rm d}}{{\rm d}t} \left( \begin{matrix} \nu_e \\ \nu_\mu \\ \nu_\tau \end{matrix} \right) = \frac{1}{2E} \left[ U \left( \begin{matrix} 0 & 0 & 0 \\ 0 & \Delta m_{21}^2 & 0 \\ 0 & 0 & \Delta m_{31}^2 \end{matrix} \right) U^\dagger + A \left( \begin{matrix} 1 + \varepsilon_{ee} & \varepsilon_{e\mu} & \varepsilon_{e\tau} \\ \varepsilon_{e\mu}^* & \varepsilon_{\mu\mu} & \varepsilon_{\mu\tau} \\ \varepsilon_{e\tau}^* & \varepsilon_{\mu\tau}^* & \varepsilon_{\tau\tau} \end{matrix} \right) \right] \left( \begin{matrix} \nu_e \\ \nu_\mu \\ \nu_\tau \end{matrix} \right) \,.
\label{eq:3prop}
\end{equation}
The `1' in the 1-1--element of the effective matter potential in equation~(\ref{eq:3prop}) describes the weak interaction of electron neutrinos with left-handed electrons through the exchange of W bosons, i.e.~the standard matter interactions, whereas the NSI parameters $\varepsilon_{\alpha\beta}$ in the effective matter potential describe the matter NSIs. See figure~\ref{fig:NSIschematic} for schematic pictures of standard and non-standard matter effects.
Now, the effective Hamiltonian $\hat{H}$ in equation~(\ref{eq:Heff}), which is Hermitian, can be diagonalized using a unitary transformation, and one finds
\begin{equation}
\hat{H} = \frac{1}{2E} \tilde{U} {\rm diag}(\tilde{m}_1^2,\tilde{m}_2^2,\tilde{m}_3^2) \tilde{U}^\dagger \,,
\end{equation}
where $\tilde{m}_i^2$ ($i = 1,2,3$) denote the effective mass-squared eigenvalues of neutrinos and $\tilde{U}$ is the effective leptonic mixing matrix in matter. Of course, all the quantities $\tilde{m}_i^2$ and $\tilde{U}$ will in general be dependent on the effective matter potential $A$ as well as some of the various matter NSI parameters $\varepsilon_{\alpha\beta}$. Explicit expressions for these quantities can be found in \cite{Meloni:2009ia}.

In the case of matter NSIs, for a constant matter density profile (which is close to reality for most long-baseline neutrino oscillation experiments), the neutrino transition probabilities are given by
\begin{equation}
P(\nu_\alpha \to \nu_\beta;L) = \left| \sum_{i=1}^3 \tilde{U}_{\alpha i} \tilde{U}_{\beta i}^* \, {\rm e}^{-{\rm i} \frac{\tilde{m}_i^2 L}{2E}} \right|^2 \,,
\label{eq:Pm}
\end{equation}
where $L$ is the baseline length. Comparing equation~(\ref{eq:Pm}) with the formula for neutrino transition probabilities in vacuum (i.e.~equation~(\ref{eq:3nu})), one arrives at the conclusion that there is no difference between the form of the neutrino transition probabilities in matter with NSIs and in vacuum if one replaces the effective parameters $\tilde{m}_i^2$ and $\tilde{U}$ in equation~(\ref{eq:Pm}) by the vacuum parameters $m_i^2$ and $U$. The mappings between the effective parameters and the vacuum ones are sufficient to study new physics effects entering future long-baseline neutrino oscillation experiments (see section~\ref{sub:other_scenarios}). The important point is the dia\-go\-na\-li\-za\-tion of the effective Hamiltonian $\hat{H}$ and the derivation of the explicit expressions for the effective parameters. Now, using equation~(\ref{eq:Pm}), we can express the neutrino oscillation probabilities in matter with NSIs (for a realistic experiment) as follows \cite{Meloni:2009ia}:
\begin{eqnarray}\label{eq:PJ1}
P(\nu_\alpha \to \nu_\alpha;L) & = & 1-4\sum_{i>j} |\tilde U_{\alpha i} \tilde
U^*_{\alpha j}|^2 \sin^2\left(\frac{\Delta \tilde{m}_{ij}^2 L}{4E}\right) \,, \\
P(\nu_\alpha \to \nu_\beta;L) & = &  - 4\sum_{i>j} {\rm Re}\left( \tilde
U^*_{\alpha i} \tilde U_{\beta i}\tilde U_{\alpha j}\tilde
U^*_{\beta j}\right)\sin^2\left(\frac{\Delta \tilde{m}_{ij}^2 L}{4E}\right) - 8 {\cal J} \prod_{i>j}
\sin\left(\frac{\Delta \tilde{m}_{ij}^2 L}{4E}\right) \,, \label{eq:PJ2}
\end{eqnarray}
where $(\alpha,\beta)$ run over $(e,\mu)$, $(\mu,\tau)$ and $(\tau,e)$ and the quantity ${\cal J}$ is defined through the relation
\begin{eqnarray}\label{eq:Jm}
{\cal J}^2 & = & |\tilde{U}_{\alpha i}|^2 |\tilde{U}_{\beta
j}|^2 |\tilde{U}_{\alpha j}|^2 |\tilde{U}_{\beta i} |^2 -
\frac{1}{4} \left (1 + |\tilde{U}_{\alpha i}|^2 |\tilde{U}_{\beta
j}|^2 + |\tilde{U}_{\alpha j}|^2 |\tilde{U}_{\beta i}|^2 \right .
\nonumber \\
& & \left . - |\tilde{U}_{\alpha i}|^2 - |\tilde{U}_{\beta j}|^2 -
|\tilde{U}_{\alpha j}|^2 - |\tilde{U}_{\beta i}|^2 \right )^2 \,.
\end{eqnarray}

\subsection{Mappings with matter NSIs}

\begin{figure}
  \includegraphics[width=0.95\textwidth,bb = 30 350 530 650]{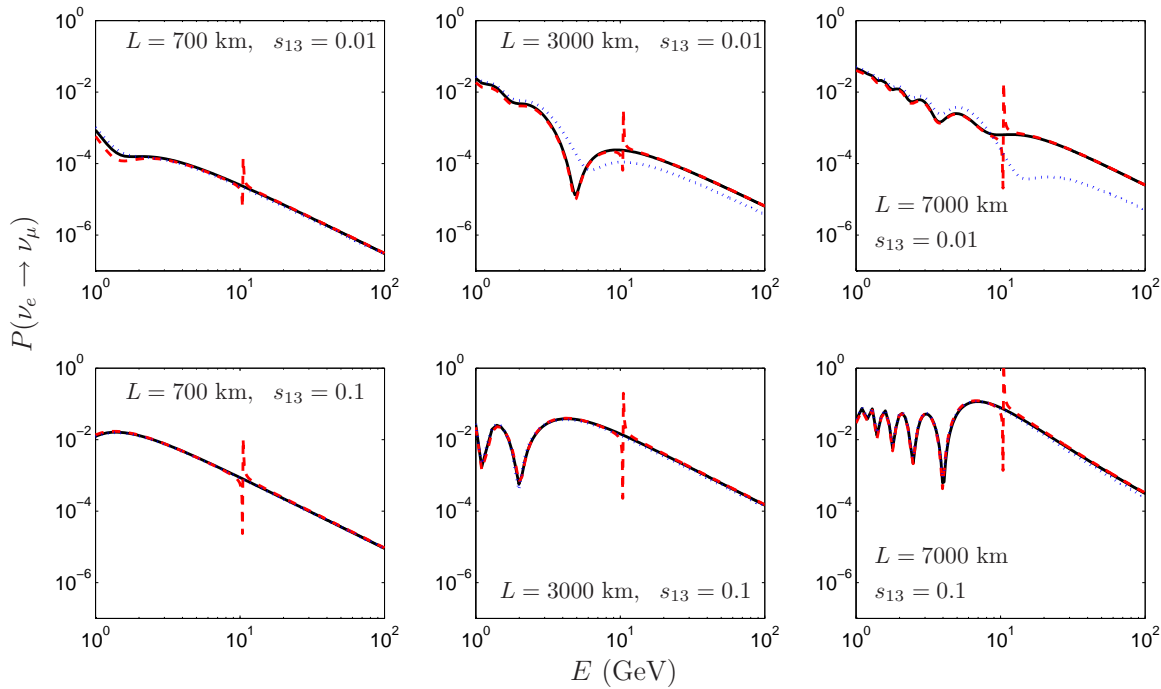}
  \vspace{-5mm}
  \caption{Neutrino oscillation probabilities for the $\nu_e \to \nu_\mu$ channel as functions of the neutrino energy $E$. We have set $\delta = \pi/2$ and $\varepsilon_{e\tau} = 0.01$, and all other matter NSI parameters are zero. Solid (black) curves are exact nu\-me\-ri\-cal results, dashed (red) curves are the approximative results and dotted (blue) curves are results without NSIs. This figure has been reproduced with permission from \protect\cite{Meloni:2009ia}.}
  \label{fig:prob}
\end{figure}

In \cite{Meloni:2009ia}, using first-order non-degenerate perturbation theory in the mass hierarchy parameter $\alpha \equiv \Delta m_{21}^2/\Delta m_{31}^2$, the smallest leptonic mixing angle $s_{13} \equiv \sin \theta_{13}$ and all the matter NSI parameters $\varepsilon_{\alpha\beta}$, model-independent mappings for the effective masses with NSIs during propagation processes, i.e.~in matter, were derived, which are given by
\begin{eqnarray}\label{eq:approxm}
\tilde m^2_1 & \simeq & \Delta m^2_{31} \left(\hat A + \alpha s^2_{12} +
\hat A\varepsilon_{ee}\right) \,, \\
\tilde m^2_2 & \simeq & \Delta m^2_{31} \left[ \alpha c^2_{12} - \hat A
s^2_{23} \left(\varepsilon_{\mu\mu}-\varepsilon_{\tau\tau}\right) -
\hat A s_{23} c_{23} \left(\varepsilon_{\mu\tau} +
\varepsilon^*_{\mu\tau}\right) + \hat A \varepsilon_{\mu\mu} \right] \,, \label{eq:approxm2}\\
\tilde m^2_3 & \simeq & \Delta m^2_{31} \left[ 1 + \hat A
\varepsilon_{\tau\tau} + \hat A s^2_{23}
\left(\varepsilon_{\mu\mu}-\varepsilon_{\tau\tau}\right) + \hat A
s_{23} c_{23} \left(\varepsilon_{\mu\tau} +
\varepsilon^*_{\mu\tau}\right)\right] \,,
\end{eqnarray}
as well as model-independent mappings for the effective mixing matrix
elements with NSIs, which are given by
\begin{eqnarray}
\tilde U_{e2} & \simeq & \frac{\alpha s_{12} c_{12}}{\hat A} +
c_{23}\varepsilon_{e\mu} - s_{23} \varepsilon_{e\tau}  \,, \label{eq:approxV2}\\
\tilde U_{e3} & \simeq &  \frac{s_{13}{\rm e}^{-{\rm i}\delta}}{1-\hat A}
+ \frac{\hat A
(s_{23}\varepsilon_{e\mu}+ c_{23}\varepsilon_{e\tau})  }{1-\hat A} \,, \label{eq:approxV1}\\
\tilde U_{\mu 2} & \simeq & c_{23} + \hat{A} s_{23}^2 c_{23} \left( \varepsilon_{\tau\tau} - \varepsilon_{\mu\mu} \right) + \hat{A} s_{23} \left( s_{23} \varepsilon_{\mu\tau} - c_{23}^2 \varepsilon_{\mu\tau}^* \right) \,, \\
\tilde U_{\mu 3} & \simeq & s_{23} + \hat A  \left[ c_{23} \varepsilon_{\mu\tau}
+ s_{23} c_{23}^2 \left(  \varepsilon_{\mu\mu} - \varepsilon_{\tau\tau} \right)
 - s^2_{23} c_{23}  \left(
\varepsilon_{\mu\tau} + \varepsilon^*_{\mu\tau} \right)
\right] \,, \label{eq:approxV3}
\end{eqnarray}
where $\hat{A} \equiv A/\Delta m^2_{31}$. In equation~(\ref{eq:approxV2}), there is an unphysical divergence for $\hat{A} \to 0$, whereas in equation~(\ref{eq:approxV1}), there is a resonance at $\hat{A} = 1$, which are both well-known consequences of non-degenerate perturbation theory. Thus, degenerate perturbation theory needs to be used around these two singularities. Note that equations~(\ref{eq:approxm})--(\ref{eq:approxV3}) are first-order series expansions in the small parameters $\alpha$, $s_{13}$ and $\varepsilon_{\alpha\beta}$, i.e.~linear in these parameters, but valid to all orders in all other parameters. Furthermore, note that only equation~(\ref{eq:approxV1}) is explicitly linearly dependent on $s_{13}$, only equations~(\ref{eq:approxm}), (\ref{eq:approxm2}) and (\ref{eq:approxV2}) are explicitly linearly dependent on $\alpha$, and all equations are at least linearly dependent on one of the $\varepsilon_{\alpha\beta}$'s. We observe from the explicit mappings~(\ref{eq:approxm})--(\ref{eq:approxV3}) that the effective parameters can be totally different from the fundamental parameters, because of the dependence on $\hat{A}$ and the NSI parameters $\varepsilon_{\alpha\beta}$. In addition, in figure~\ref{fig:prob}, neutrino oscillation probabilities including the effects of NSIs for the electron neutrino-muon neutrino channel are plotted.
It is found that the approximate mappings agree with the exact numerical results to an extremely good precision. However, note that a singularity exists around 10 GeV, which corresponds to the resonance at $\hat{A} \sim 1$ and is due to the breakdown of non-degenerate perturbation theory that has been adopted to derive the model-independent mappings. Thus, the approximate model-independent mapping~(\ref{eq:approxV1}) is not valid around 10 GeV.

\subsection{Approximate formulas for two neutrino flavors with matter NSIs}
\label{sub:approx2fl}

The three-flavor neutrino evolution given in equation~(\ref{eq:3prop}) is rather complicated and cumbersome. Hence, in order to illuminate neutrino oscillations with matter NSIs, we investigate the oscillations using two flavors, e.g.~$\nu_e$ and $\nu_\tau$. In this case, we have the much simpler two-flavor neutrino evolution equation
\begin{equation}
{\rm i} \frac{{\rm d}}{{\rm d} L} \left( \begin{matrix} \nu_e \\ \nu_\tau \end{matrix} \right) = \frac{1}{2E} \left[ U \left( \begin{matrix} 0 & 0 \\ 0 & \Delta m^2 \end{matrix} \right) U^\dagger + A \left( \begin{matrix} 1 + \varepsilon_{ee} & \varepsilon_{e\tau} \\ \varepsilon_{e\tau} & \varepsilon_{\tau\tau} \end{matrix} \right) \right] \left( \begin{matrix} \nu_e \\ \nu_\tau \end{matrix} \right) \,,
\label{eq:2prop}
\end{equation}
where $L$ is the neutrino propagation length that has replaced time in equation~(\ref{eq:3prop}). Using equation~(\ref{eq:2prop}), one can derive the two-flavor neutrino oscillation probability (see equation~(\ref{eq:2nu}))
\begin{equation}
P(\nu_e \to \nu_\tau;L) = \sin^2 \left(2\tilde\theta\right) \sin^2 \left(\frac{\Delta \tilde{m}^2 L}{2E}\right) \,,
\end{equation}
where $\tilde\theta$ and $\Delta \tilde{m}^2$ are the effective neutrino oscillation parameters when taking into account matter NSIs. These parameters are related to the vacuum neutrino oscillation parameters $\theta$ and $\Delta m^2$ and given by (see, e.g., \cite{Kitazawa:2006iq})
\begin{eqnarray}
\left( \Delta \tilde{m}^2 \right)^2 &=& \left[ \Delta m^2 \cos (2\theta) - A \left(1 + \varepsilon_{ee} - \varepsilon_{\tau\tau} \right) \right]^2 + \left[ \Delta m^2 \sin (2\theta) + 2 A \varepsilon_{e\tau} \right]^2 \,, \label{eq:MSW_NSI_1}\\
\sin \left(2\tilde\theta\right) &=& \frac{\Delta m^2 \sin (2\theta) + 2A \varepsilon_{e\tau}}{\Delta \tilde{m}^2} \,. \label{eq:MSW_NSI_2}
\end{eqnarray}
In the limit $\varepsilon_{ee},\varepsilon_{e\tau},\varepsilon_{\tau\tau} \to 0$, i.e.~when NSIs vanish, equations~(\ref{eq:MSW_NSI_1}) and (\ref{eq:MSW_NSI_2}) reduce to
\begin{eqnarray}
\left( \Delta \tilde{m}_0^2 \right)^2 &=& \left[ \Delta m^2 \cos (2\theta) - A \right]^2 + \left[ \Delta m^2 \sin (2\theta) \right]^2 \,, \\
\sin \left(2\tilde\theta_0\right) &=& \frac{\Delta m^2 \sin (2\theta)}{\Delta \tilde{m}_0^2} \,,
\end{eqnarray}
which are the formulas for the ordinary MSW effect \cite{Wolfenstein:1977ue,Mikheev:1986gs,Mikheev:1986wj}. Thus, NSIs give rise to modified (and more general) versions of the MSW effect, i.e.~equations~(\ref{eq:MSW_NSI_1}) and (\ref{eq:MSW_NSI_2}). Cf.~discussion in section~\ref{sub:NSI_H_eff}. For further discussion on approximate formulae for two neutrino flavors with NSIs, see \cite{Blennow:2008eb}.

\section{Theoretical models for NSIs}
\label{sec:NSImodels}

In order to realize NSIs in a more fundamental framework with some underlying high-energy physics theory, it is generally desirable that it respects and encompasses the SM gauge group ${\rm SU(3)} \times {\rm SU(2)} \times {\rm U(1)}$. Note that the theoretical models presented in this section only represent a small selection, there exists many other models in the literature. In a toy model, including the SM and one heavy SU(2) singlet scalar field $S$ with hypercharge $-1$, we can have the following interaction Lagrangian \cite{Bilenky:1993bt}
\begin{equation}
{\cal L}_{\rm int}^S = - \lambda_{\alpha\beta} \overline{L_\alpha^c} {\rm i} \sigma_2 L_\beta S + \mbox{h.c.} \,,
\end{equation}
where the quantities $\lambda_{\alpha\beta}$ ($\alpha,\beta = e,\mu,\tau$) are elements of the asymmetric coupling matrix $\lambda$, $L_\alpha$ is a doublet lepton field and $\sigma_2$ is the second Pauli matrix. Now, integrating out the heavy field $S$, generates an anti-symmetric dimension-6 operator at tree level \cite{Antusch:2008tz}, i.e.
\begin{equation}
{\cal L}_{\rm NSI}^{\rm d=6, as} = 4 \frac{\lambda_{\alpha\beta} \lambda_{\delta\gamma}^*}{m_S^2} \left( \overline{\ell^c}_\alpha P_L \nu_\beta \right) \left( \bar\nu_\gamma P_R \ell^c_\delta \right) \,, 
\end{equation}
where $m_S$ is the mass of the heavy field $S$, while $P_L$ and $P_R$ are left- and right-handed projection operators, respectively. Note that this is the only gauge-invariant dimension-6 operator, which does not give rise to charged-lepton NSIs\footnote{Charged-lepton NSIs are non-standard inteactions originating from processes that involve charged leptons (e.g.~lepton flavor violating decays $\ell_\alpha^\mp \to \ell_\beta^\pm \ell_\gamma^\mp \ell_\delta^\mp$). See figures~\protect\ref{fig:tree} and \protect\ref{fig:diagrams}.}.

\subsection{Gauge symmetry invariance}

At high-energy scales, where NSIs originate, there exists ${\rm SU(2)} \times {\rm U(1)}$ gauge symmetry invariance. In general, theories beyond the SM must respect gauge symmetry invariance, which implies strict constraints on possible models for NSIs (see, e.g., \cite{Gavela:2008ra}). Therefore, if there is a dimension-6 operator on the form
$$
\frac{1}{\Lambda^2} \left( \bar\nu_\alpha \gamma^\rho P_L \nu_\beta \right) \left( \bar\ell_\gamma \gamma_\rho P_L \ell_\delta \right) \,,
$$
then this operator will lead to NSI parameters such as $\varepsilon_{e\mu}^{ee} \, (=\varepsilon_{e\mu}^{eeL})$. However, the above form for a dimension-6 operator must be a part of the more general form
$$
\frac{1}{\Lambda^2} \left( \bar L_\alpha \gamma^\rho L_\beta \right) \left( \bar L_\gamma \gamma_\rho L_\delta \right) \,,
$$
which involves four charged-lepton operators. Thus, we have severe constraints from experi\-ments on processes like $\mu \to 3 e$,\footnote{In general, both lepton flavor violating process (such as $\mu \to 3 e$) and allowed regions for fundamental neutrino parameters (such as neutrino mass-squared differences and leptonic mixing angles) set constraints on NSIs, but normally lepton flavor violating processes put stronger bounds on the NSIs than the fundamental neutrino parameters. See, e.g., the discussions in sections~\protect\ref{sub:tsm} and \protect\ref{sub:ZBm}.} i.e.
$$
{\rm BR}(\mu \to 3e) < 10^{-12} \,,
$$
which leads to the following upper bound on the above chosen NSI parameter
$$
\varepsilon_{e\mu}^{ee} < 10^{-6} \,.
$$
Note that the above discussion is only valid for dimension-6 operators, and can be extended to operators with dimension equal to 8 or larger, but this will not be performed here.

\subsection{A seesaw model---the triplet seesaw model}
\label{sub:tsm}

\begin{figure}
  \includegraphics[height=.225\textheight]{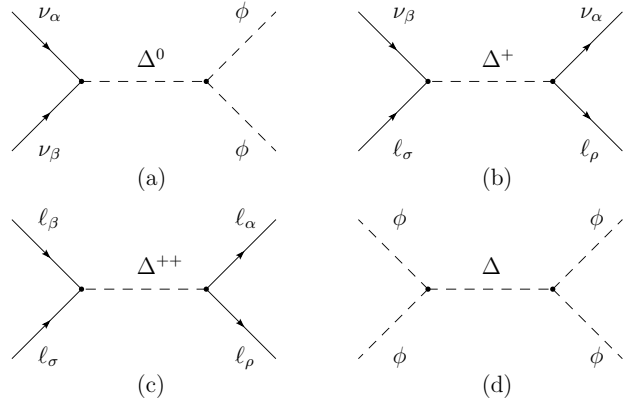}
  \vspace{-2.5mm}
  \caption{Tree-level diagrams with exchange of a heavy Higgs triplet in the triplet seesaw model. This figure has been reproduced with permission from \cite{Malinsky:2008qn}. Copyright (2009) by The American Physical Society.}
  \label{fig:tree}
\end{figure}

In a type-II seesaw model (also known as the triplet seesaw model), the tree-level diagrams with exchange of a heavy Higgs triplet are given in figure~\ref{fig:tree}.
Integrating out the heavy triplet field (at tree level), we obtain the relations between the NSI parameters and the elements of the light neutrino mass matrix as \cite{Malinsky:2008qn}
\begin{equation}
\varepsilon_{\alpha\beta}^{\rho\sigma} = - \frac{m_\Delta^2}{8 \sqrt{2} G_F v^4 \lambda_\phi^2} (m_\nu)_{\sigma\beta} \left( m_\nu^\dagger \right)_{\alpha\rho} \,,
\label{eq:epsilon=mm}
\end{equation}
where $v \simeq 174$~GeV is the vacuum expectation value of the SM Higgs field\footnote{The vacuum expectation value of the SM Higgs field is normally defined as $v = \left( \sqrt{2} G_F \right)^{-1/2} \simeq 246 \, {\rm GeV}$. Thus, the two values 174 GeV and 246 GeV differ by a factor $\sqrt{2}$.}, $m_\Delta$ is the mass of the Higgs triplet field and $\lambda_\phi$ is associated with the trilinear Higgs coupling. It holds that the absolute neutrino mass scale is proportional to $\lambda_\phi v^2/m_\Delta^2$, which means that $\left( m_\nu \right)_{\alpha\beta} \sim \lambda_\phi v^2/m_\Delta^2$. Thus, inserting the proportionality of the elements of the light neutrino mass matrix into equation~(\ref{eq:epsilon=mm}) and using the relation $G_F/\sqrt{2} \simeq g_W^2/(8 m_W^2)$, we find that
\begin{equation}
\varepsilon_{\alpha\beta}^{\rho\sigma} \propto \frac{m_\Delta^2}{\frac{g_W^2}{m_W^2} \cdot v^2 \lambda_\phi^2} \cdot \frac{\lambda_\phi v^2}{m_\Delta^2} \cdot \frac{\lambda_\phi v^2}{m_\Delta^2} = \frac{m_W^2}{m_\Delta^2} \,,
\end{equation}
which has the characteristic dependence given in equation~(\ref{eq:mW2mX2}).

Now, using experimental constraints from lepton flavor violating processes (rare lepton decays and muonium-antimuonium conversion) \cite{Willmann:1998gd,Beringer:1900zz}, we find upper bounds on the NSI parameters, which are presented in table~\ref{tab:bounds}. From this table, we can observe that the NSI parameter $\varepsilon_{\mu e}^{\mu e}$ has the weakest upper bound.

\begin{table}
\begin{tabular}{lcc}
\hline
Decay & Constraint on & Bound
\\
\hline
$\mu^- \rightarrow e^- e^+ e^- $  &  $| \varepsilon^{e \mu}_{ee} |$ & $ 3.5 \times 10^{-7}$  \\
$\tau^-
\rightarrow e^- e^+  e^- $  & $| \varepsilon^{e \tau}_{ee} |$  & $ 1.6 \times 10^{-4}$  \\
$\tau^- \rightarrow  \mu^- \mu^+ \mu^-$  & $| \varepsilon^{\mu \tau}_{\mu\mu} |$  & $ 1.5 \times 10^{-4}$  \\
$\tau^- \rightarrow e^- \mu^+  e^-$  & $| \varepsilon^{e \tau}_{e\mu} |$  & $1.2 \times 10^{-4}$  \\
$\tau^- \rightarrow \mu^- e^+  \mu^-$  & $| \varepsilon^{\mu \tau}_{\mu e } |$  &  $ 1.3 \times 10^{-4}$  \\
$\tau^- \rightarrow e^- \mu^+ \mu^-  $  & $| \varepsilon^{e \tau}_{\mu\mu} |$  &  $ 1.2 \times 10^{-4}$ \\
$\tau^- \rightarrow e^- e^+ \mu^-$  & $| \varepsilon^{e \tau}_{\mu e} |$  &  $ 9.9 \times 10^{-5}$ \\
$\mu^- \rightarrow e^- \gamma  $  &  $| \sum_{\alpha} \varepsilon^{e \mu}_{\alpha\alpha} |$ &  $ 1.4 \times 10^{-4}$ \\
$\tau^- \rightarrow e^- \gamma  $  & $| \sum_{\alpha} \varepsilon^{e \tau}_{\alpha\alpha} |$  &  $3.2 \times 10^{-2}$ \\
$\tau^- \rightarrow \mu^- \gamma  $  & $| \sum_{\alpha} \varepsilon^{\mu \tau}_{\alpha\alpha} |$  &  $2.5 \times 10^{-2}$ \\
$\mu^+ e^- \rightarrow \mu^- e^+$  &  $| \varepsilon^{\mu e}_{\mu e} |$  &  $ 3.0 \times 10^{-3}$ \\
\hline
\end{tabular}
\caption{Constraints on various NSI parameters from $\ell \rightarrow \ell\ell\ell$, one-loop $\ell \rightarrow \ell \gamma$ and $\mu^+ e^- \rightarrow \mu^- e^+$ processes. Copyright (2009) by The American Physical Society.} \vspace{-0.5cm}
\label{tab:bounds}
\end{table}

\begin{figure}
  \includegraphics[height=0.4\textheight]{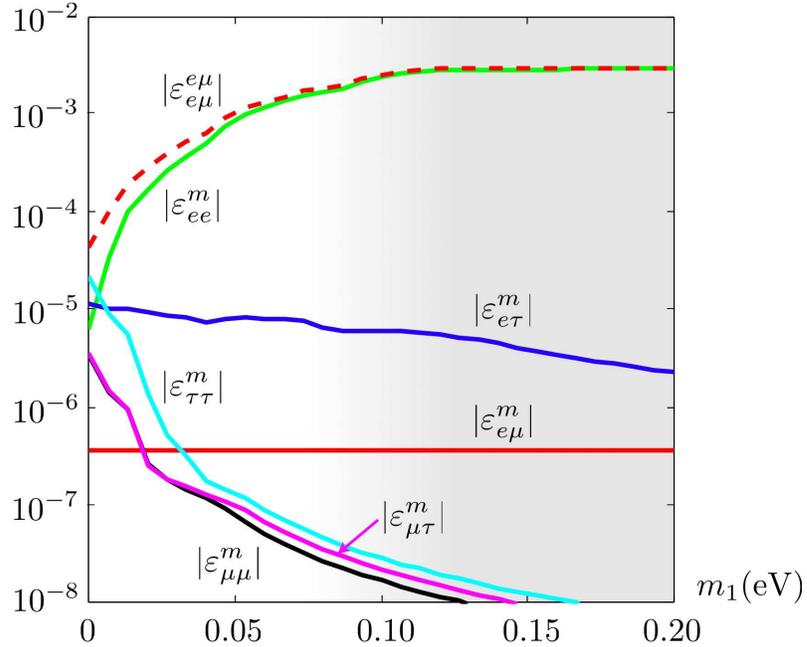}
  \vspace{-5mm}
  \caption{Upper bounds on various NSI parameters in the triplet seesaw model. Note that the matter NSI parameters are defined as $\varepsilon_{\alpha\beta}^m \equiv \varepsilon_{\alpha\beta}^{ee}$. This figure has been reproduced with permission from \cite{Malinsky:2008qn}. Copyright (2009) by The American Physical Society.}
  \label{fig:NSIs}
\end{figure}

In addition, for $m_\Delta = 1 \, {\rm TeV}$, using the constraints on lepton flavor violating processes, and varying $m_1$, we plot the upper bounds on some of the NSI parameters in the triplet seesaw model. The results are shown in figure~\ref{fig:NSIs}.
For a hierarchical mass spectrum (i.e.~$m_1 < 0.05 \, {\rm eV}$), all the NSI effects are suppressed, whereas for a nearly degenerate mass spectrum (i.e.~$m_1 > 0.1 \, {\rm eV}$), two NSI parameters can be sizable, which are $\varepsilon_{e\mu}^{e\mu}$ and $\varepsilon_{ee}^m \equiv \varepsilon_{ee}^{ee}$.

\subsection{The Zee--Babu model}
\label{sub:ZBm}

\begin{figure}
  \vspace{0.cm}
  \includegraphics[width=0.575\textwidth]{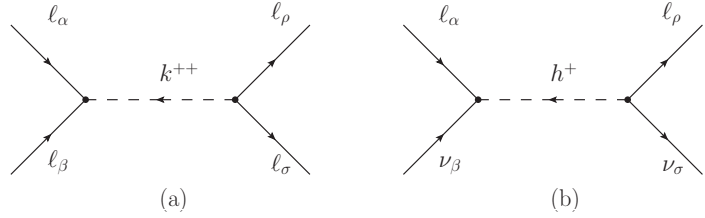}
  \vspace{-2.5mm}
  \caption{Tree-level diagrams for the exchange of heavy scalars in the Zee--Babu model. This figure is an updated and corrected version of one of the figures from \protect\cite{Ohlsson:2009vk}.}
  \label{fig:diagrams}
  \vspace{-0.cm}
\end{figure}

In the Zee--Babu model \cite{Zee:1985rj,Zee:1985id,Babu:1988ki}, we have the Lagrangian
\begin{equation}
{\cal L} = {\cal L}_{\rm SM} + f_{\alpha\beta} L_{{\rm L}\alpha}^T C {\rm i} \sigma_2 L_{{\rm L}\beta} h^+ + g_{\alpha\beta} \overline{e_\alpha^c} e_\beta k^{++} - \mu h^- h^- k^{++} + {\rm h.c.} + V_H \,,
\end{equation}
where $f_{\alpha\beta}$ and $g_{\alpha\beta}$ are antisymmetric and symmetric Yukawa couplings, respectively, and $h^+$ and $k^{++}$ are heavy charged scalars that could be observed at the LHC and which lead to a two-loop diagram that generates small neutrino masses. The tree-level diagrams that are responsible for (a) non-standard interactions of four charged leptons and (b) NSIs (neutrinos) are presented in figure~\ref{fig:diagrams}.
Using these diagrams, both types of non-standard lepton interactions are obtained after integrating out the heavy scalars, which induces three relevant (and potentially sizable) matter NSI parameters ($\varepsilon_{\mu\tau}^m$, $\varepsilon_{\mu\mu}^m$ and $\varepsilon_{\tau\tau}^m$) and one production NSI parameter ($\varepsilon_{\mu\tau}^s$, which is important for the $\nu_\mu \to \nu_\tau$ channel at a future neutrino factory, see also section~\ref{sub:nf}) that are given by
\begin{eqnarray}
\varepsilon_{\alpha\beta}^m &=& \varepsilon_{\alpha\beta}^{ee} = \frac{f_{e\beta} f_{e\alpha}^*}{\sqrt{2} G_F m_h^2} \simeq \frac{4 f_{e\beta} f_{e\alpha}^*}{g_W^2} \frac{m_W^2}{m_h^2} \propto \frac{m_W^2}{m_h^2} \,, \\
\varepsilon_{\mu\tau}^s &=& \varepsilon_{\tau e}^{e\mu} = \frac{f_{\mu e} f_{e\tau}^*}{\sqrt{2} G_F m_h^2} \simeq \frac{4 f_{\mu e} f_{e\tau}^*}{g_W^2} \frac{m_W^2}{m_h^2} \propto \frac{m_W^2}{m_h^2} \,,
\end{eqnarray}
where we observe that the NSI parameters in the Zee--Babu model also have the characteristic dependence given in equation~(\ref{eq:mW2mX2}), which means that they naively are in the range $10^{-4} - 10^{-2}$ if the scale of the heavy scalar masses is of the order of $(1 - 10) \, {\rm TeV}$.

\begin{figure}
  \vspace{-1cm}
  \includegraphics[width=7.5cm,bb=0 0 750 750]{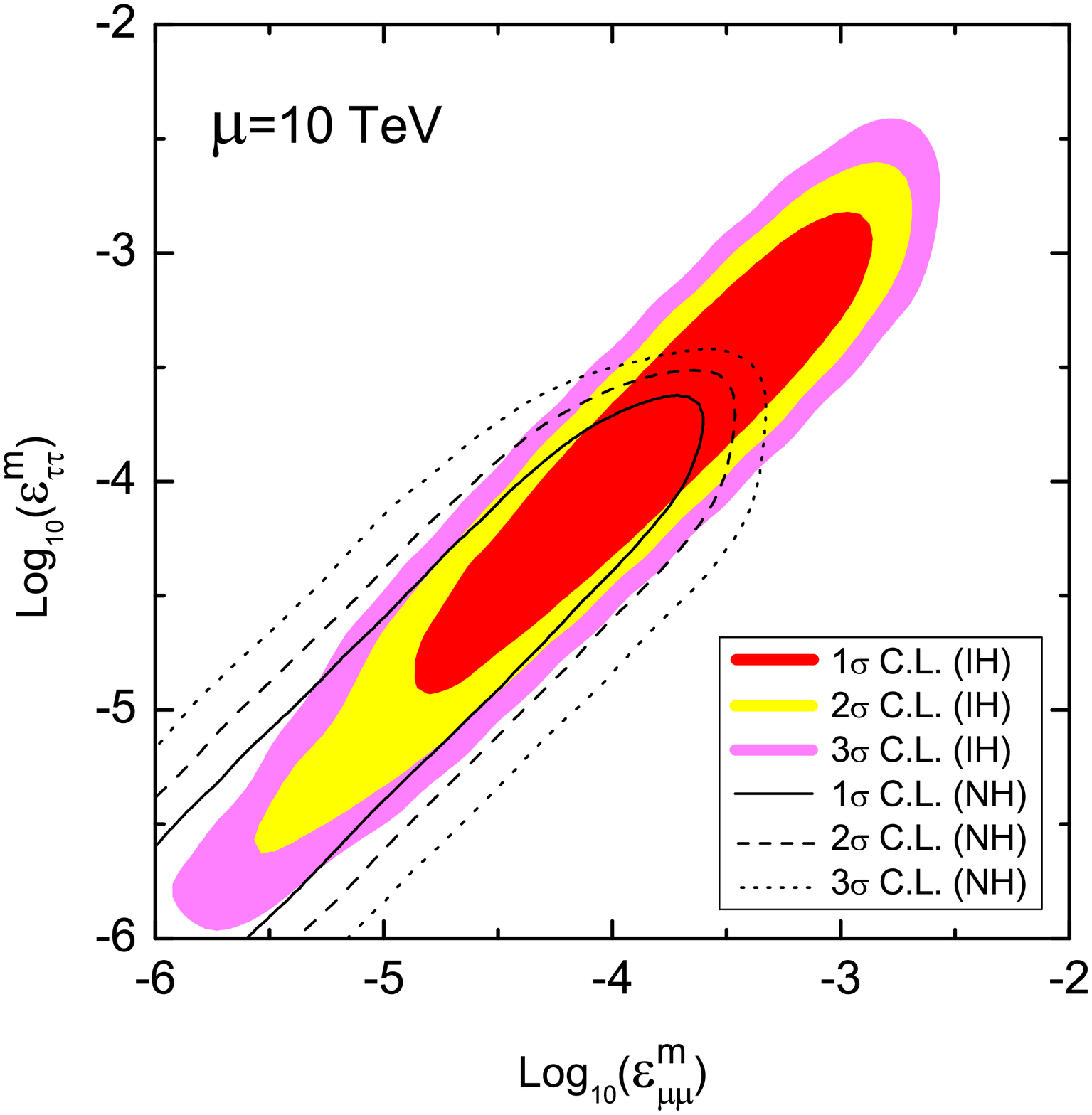}
  \includegraphics[width=7.5cm,bb=0 0 750 750]{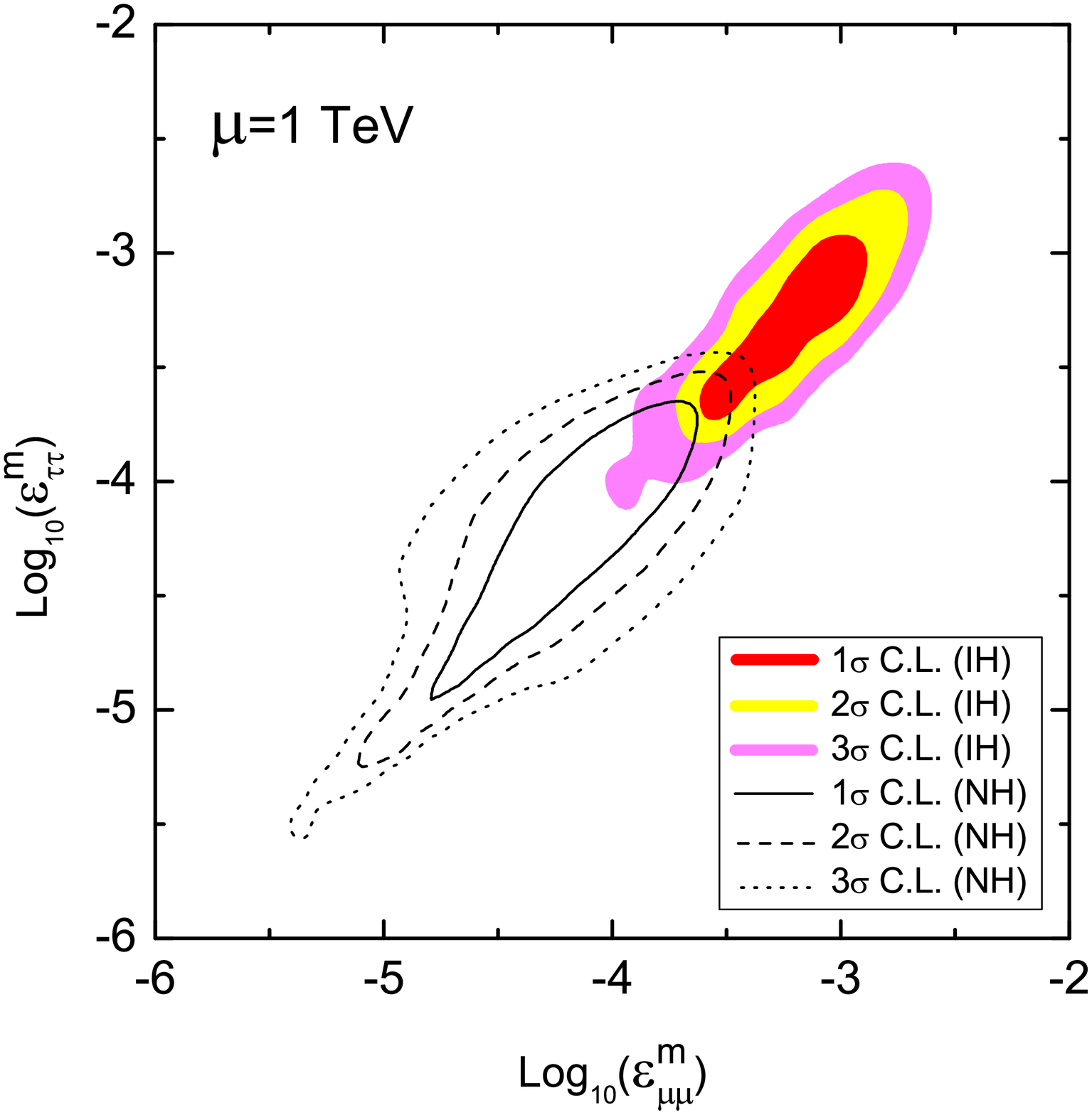}
  \vspace{-0.25cm}
  \caption{Allowed region of NSI parameters $\varepsilon_{\mu\mu}^m$ and $\varepsilon_{\tau\tau}^m$ at 1$\sigma$, 2$\sigma$ and 3$\sigma$ confidence level (C.L.) in the Zee--Babu model. The following values have been used for the heavy scalar masses: $m_h = m_k = \mu = 10$~TeV for the left plot and $m_h = m_k = \mu = 1$~TeV for the right plot. This figure has been reproduced with permission from \protect\cite{Ohlsson:2009vk}.}
  \label{fig:zeebabu}
\end{figure}

In figure~\ref{fig:zeebabu}, using best-fit values of the neutrino mass-squared differences (while taking the leptonic mixing angles to be independent parameters) \cite{Schwetz:2008er} and experimental constraints on lepton flavor violating processes (such as rare lepton decays and muonium--antimuonium conversion) \cite{Nebot:2007bc}, the allowed regions of the matter NSI parameters $\varepsilon_{\mu\mu}^m$ and $\varepsilon_{\tau\tau}^m$ in the Zee--Babu model are plotted for heavy scalar masses of 10~TeV (left plot) and 1~TeV (right plot).
Indeed, since the leptonic mixing angles are free parameters with constraints (taken from \cite{Schwetz:2008er}), their allowed regions can change when the values of the NSI parameters become non-zero. In the case of inverted neutrino mass hierarchy, the matter NSI parameters $\varepsilon_{\mu\mu}^m$ and $\varepsilon_{\tau\tau}^m$ could be in the range $10^{-4} - 10^{-3}$, whereas in the case of normal neutrino mass hierarchy, they are normally at least one order of magnitude smaller \cite{Ohlsson:2009vk}. Note that the size of $\varepsilon_{\mu\mu}^m$ and $\varepsilon_{\tau\tau}^m$ may be too small to be observable, whereas $\varepsilon_{\mu\tau}^m$ could be within the reach of a future neutrino factory. In addition, for inverted neutrino mass hierarchy and heavy scalar masses of 1~TeV, it turns out that $\varepsilon_{\mu\tau}^s$ is predicted to also be in the range $10^{-4} - 10^{-3}$, which is probably in the reach for a near tau-detector at a future neutrino factory \cite{Ohlsson:2009vk}.

\section{Phenomenology of NSIs for different types of experiments}
\label{sec:NSIpheno}

In this section, we will discuss the phenomenology of NSIs for atmospheric, accelerator and reactor neutrino experiments as well as for neutrino factory setups and astrophysical settings such as solar and supernova neutrinos. As we will see, only two experimental collaborations have used their neutrino data to analyze NSIs, which are the Super-Kamiokande and MINOS collaborations.

\subsection{Atmospheric neutrino experiments}
\label{sub:atmosne}

Neutrino oscillations with matter NSIs that are important for atmospheric neutrinos have been previously studied in the literature. For example, there are phenomenological studies that investigate the possibility to probe NSIs with atmospheric neutrino data only \cite{Fornengo:2001pm,GonzalezGarcia:2004wg,Friedland:2004ah} and in combination with other neutrino data \cite{Huber:2001zw,Guzzo:2001mi,Friedland:2005vy,GonzalezGarcia:2011my,Kopp:2010qt,Escrihuela:2011cf}, whereas there exist also more theoretical investigations \cite{Blennow:2008eb,Mann:2010jz}. However, most importantly, there is an experimental study on matter NSIs with atmospheric neutrino data from the Super-Kamiokande collaboration \cite{Mitsuka:2011ty}. In principle, atmospheric neutrinos are very sensitive to matter NSIs, since they travel over long distances inside the Earth before being detected \cite{Kopp:2010qt}.

In order to analyze atmospheric neutrino data in the simplest way, we consider a two-flavor neutrino oscillation approximation with matter NSIs in the $\nu_\mu$--$\nu_\tau$ sector (see section~\ref{sub:approx2fl}), since $\nu_\mu \leftrightarrow \nu_\tau$ oscillations are important for atmospheric neutrinos. In this case, the first row and first column in equation~(\ref{eq:3prop}) are cancelled, which effectively means that the NSI parameters that couple to $\nu_e$ are set to zero, i.e.~$\varepsilon_{e \alpha} = 0$, where $\alpha = e,\mu,\tau$. In addition, the parameters $\Delta m_{21}^2$, $\theta_{12}$ and $\theta_{13}$ are not important, leading to a two-flavor approximation that only includes the parameters $\Delta m^2 \equiv \Delta m_{31}^2$, $\theta \equiv \theta_{23}$, $\varepsilon_{\mu\mu}$, $\varepsilon_{\mu\tau} = \varepsilon_{\tau\mu}$\footnote{Note that the assumption that the off-diagonal NSI parameters are real is not generic.} and $\varepsilon_{\tau\tau}$. In this approximation, which has been named the {\it two-flavor hybrid model}, the two-flavor neutrino evolution equation reads
\begin{equation}
{\rm i} \frac{{\rm d}}{{\rm d} L} \left( \begin{matrix} \nu_\mu \\ \nu_\tau \end{matrix} \right) = \frac{1}{2E} \left[ U \left( \begin{matrix} 0 & 0 \\ 0 & \Delta m^2 \end{matrix} \right) U^\dagger + A \left( \begin{matrix} 1 + \varepsilon_{\mu\mu} & \varepsilon_{\mu\tau} \\ \varepsilon_{\mu\tau} & \varepsilon_{\tau\tau} \end{matrix} \right) \right] \left( \begin{matrix} \nu_\mu \\ \nu_\tau \end{matrix} \right) \,.
\label{eq:2propmt}
\end{equation}
Using equation~(\ref{eq:2propmt}), defining $\varepsilon \equiv \varepsilon_{\mu\tau}$ and $\varepsilon' \equiv \varepsilon_{\tau\tau} - \varepsilon_{\mu\mu}$, and assuming that neutrinos have NSIs with $d$-quarks only \cite{GonzalezGarcia:1998hj,Fornengo:2001pm}, we obtain the two-flavor $\nu_\mu$ survival probability \cite{GonzalezGarcia:2004wg,Mitsuka:2011ty}
\begin{equation}
P(\nu_\mu \to \nu_\mu;L) = 1 - P(\nu_\mu \to \nu_\tau;L) = 1 - \sin^2 (2\Theta) \sin^2 \left( \frac{\Delta m^2 L}{4E} R \right) \,,
\end{equation}
where the quantities $\Theta$ and $R$ are given by
\begin{eqnarray}
\sin^2 (2\Theta) &=& \frac{1}{R^2} \left[ \sin^2 (2\theta) + R_0^2 \sin^2 (2\xi) + 2 R_0 \sin (2\theta) \sin (2\xi) \right] \,, \\
R &=& \sqrt{1 + R_0^2 + 2 R_0 \left[ \cos (2\theta) \cos (2\xi) + \sin (2\theta) \sin (2\xi) \right]}
\end{eqnarray}
with the two auxiliary parameters
\begin{eqnarray}
R_0 &=& \sqrt{2} G_F N_e \frac{4E}{\Delta m^2} \sqrt{|\varepsilon|^2 + \frac{\varepsilon'^2}{4}} \,, \label{eq:R0}\\
\xi &=& \frac{1}{2} \arctan \left(\frac{2\varepsilon}{\varepsilon'}\right) \,.
\end{eqnarray}

\begin{figure}
  \includegraphics[width=0.5\textwidth]{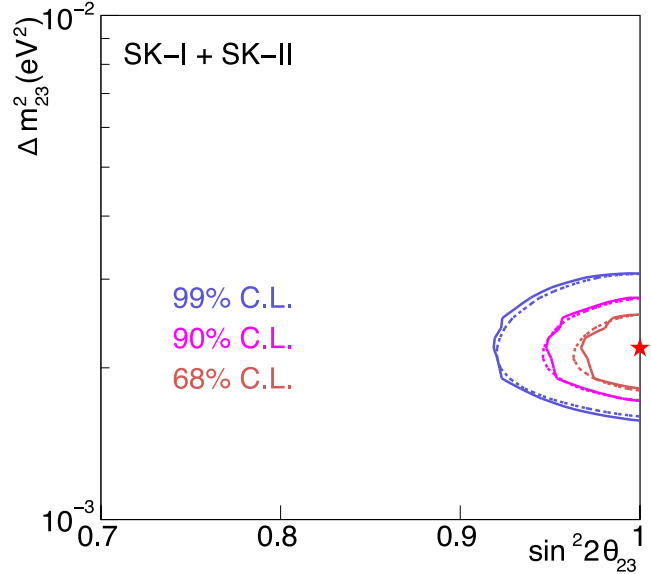}
  \vspace{-2.5mm}
  \caption{Allowed parameter regions for $\sin^2(2\theta_{23})$ and $\Delta m_{32}^2$ using the two-flavor hybrid model (solid curves) and standard two-flavor neutrino oscillations (dashed curves). The undisplayed parameters $\varepsilon$ and $\varepsilon'$ have been integrated out. This figure has been adopted from \cite{Mitsuka:2011ty}. Copyright (2011) by The American Physical Society.}
  \label{fig:SuperKNSIs}
\end{figure}

Now, using the two-flavor hybrid model together with atmospheric neutrino data from the Super-Kamiokande I (1996--2001) and II (2003--2005) experiments, the Super-Kamiokande collaboration has obtained the following results at 90\% confidence level (C.L.) \cite{Mitsuka:2011ty}
$$
|\varepsilon_{\mu\tau}| < 0.033 \quad \mbox{and} \quad |\varepsilon_{\tau\tau} - \varepsilon_{\mu\mu}| < 0.147
$$
and the allowed parameter regions for $\sin^2(2\theta_{23})$ and $\Delta m_{31}^2 \simeq \Delta m_{32}^2$ with and without NSIs are shown in figure~\ref{fig:SuperKNSIs}\footnote{It should be noted that the Super-Kamiokande (SK) collaboration uses a different convention for the NSI parameters, i.e.~$\varepsilon_{\alpha\beta}^{\rm SK} \equiv \tfrac{1}{3} \varepsilon_{\alpha\beta}$, due to the usage of the fermion number density $N_f \equiv N_d \simeq 3 N_e$ \protect\cite{GonzalezGarcia:1998hj,Fornengo:2001pm,Mitsuka:2011ty} instead of the electron number density $N_e$ in equations~(\protect\ref{eq:2propmt}) and (\protect\ref{eq:R0}), which means that the upper bounds in \protect\cite{Mitsuka:2011ty} have to be multiplied by a factor of 3.}.
In principle, there are no significant differences between the allowed parameters regions with NSIs and the ones without NSIs\footnote{Although there are no significant differences, the inclusion of NSIs in the analysis changes slightly the allowed parameter regions for the leptonic mixing angle and the neutrino mass-squared difference.}. In general, the introduction of NSIs enlarges the parameter space and enhances possible entanglements between the fundamental neutrino parameters and the NSI parameters, but since the difference between the two minimum $\chi^2$-function values (with and without NSIs) is small in the analysis of the Super-Kamiokande collaboration, no significant contribution from NSIs to ordinary two-flavor neutrino oscillations is found \cite{Mitsuka:2011ty}. Of course, this analysis can be extended to a similar analysis with a three-flavor hybrid model also taking into account the NSI parameters $\varepsilon_{ee}$ and $\varepsilon_{e\tau}$, which, however, leads to no significant changes for the allowed values of the parameter regions compared to the two-flavor hybrid model. It should be noted that the atmospheric neutrino data have no possibility to constrain the NSI parameter $\varepsilon_{ee}$ \cite{Friedland:2005vy} and the other NSI parameter $\varepsilon_{e\tau}$ is related to both $\varepsilon_{ee}$ and $\varepsilon_{\tau\tau}$ via the expression $\varepsilon_{\tau\tau} \sim |\varepsilon_{e\tau}|^2/(1 + \varepsilon_{ee})^2$ \cite{Friedland:2004ah,Friedland:2005vy,Mitsuka:2011ty}, which leads to an energy-independent parabola in the NSI parameter space spanned by $\varepsilon_{e\tau}$ and $\varepsilon_{\tau\tau}$ for a fixed value of $\varepsilon_{ee}$ and values of NSI parameters on this parabola cannot be ruled out. The reason why atmospheric neutrinos cannot constrain $\varepsilon_{ee}$ is that if $\varepsilon_{e\tau}$ is equal to zero, then the matter eigenstates are equivalent to the vacuum eigenstates \cite{Mitsuka:2011ty}. Therefore, the matter eigenvalues are independent of $\varepsilon_{ee}$ and the three-flavor model reduces to two-flavor $\nu_\mu \leftrightarrow \nu_\tau$ oscillations in matter and NSIs including $\varepsilon_{\tau\tau}$ only (see equation~(\ref{eq:Pmutau})--(\ref{eq:xi})). In conclusion, the Super-Kamiokande collaboration has found no evidence for matter NSIs in its atmospheric neutrino data.

\subsection{Accelerator neutrino experiments}
\label{sub:accne}

Studies of previous, present and future setups of accelerator neutrino oscillation experiments including matter NSIs have been thoroughly investigated in the literature, especially setups with long-baselines belong to these studies. Such studies include searches for matter NSIs with the K2K experiment \cite{Friedland:2005vy}, the MINOS experiment \cite{Kitazawa:2006iq,Friedland:2006pi,Blennow:2007pu,Mann:2010jz,Kopp:2010qt,Isvan:2011fa}, the MINOS and T2K experiments \cite{Coelho:2012bp} and the OPERA experiment \cite{Ota:2002na,EstebanPretel:2008qi,Blennow:2008ym}, as well as sensitivity analyses of the NO$\nu$A experiment \cite{Friedland:2012tq} and the T2K and T2KK experiments \cite{Oki:2010uc,Yasuda:2010hw,Adhikari:2012vc}. The prospects for detecting NSIs at the MiniBooNE experiment, which has a shorter baseline, has been investigated too \cite{Johnson:2006rs,Akhmedov:2010vy}. There are also studies that are more general in character \cite{Ota:2001pw,Ribeiro:2007jq,Escrihuela:2011cf}. In addition, the MINOS experiment has recently presented the results of a search for matter NSI in form of a poster at the `Neutrino 2012' conference in Kyoto, Japan \cite{Coelho:2012}.

Using three-flavor neutrino oscillations with matter NSIs for accelerator neutrinos, we will present the important NSI parameters and flavor transition probabilities for two experiments, which are (i) the MINOS experiment with baseline length $L \simeq 735$~km (from Fermilab in Illinois, USA to Soudan mine in Minnesota, USA) and neutrino energy $E$ in the interval $(1 - 6)$~GeV and (ii) the OPERA experiment with baseline length $L \simeq 732$~km (from CERN in Geneva, Switzerland to LNGS in Gran Sasso, Italy) and average neutrino energy $E \simeq 17$~GeV. Note that the baseline lengths of the two experiments are nearly the same, but there is a difference in the neutrino energy, which is about one order of magnitude.

First, in the case of the MINOS experiment, the important NSI parameters are $\varepsilon_{e\tau}$  and $\varepsilon_{\tau\tau}$ (see equation~(\ref{eq:3prop})) and the interesting transition probability is the $\nu_\mu$ survival probability (or equivalently the $\nu_\mu$ disappearance probability), which to leading order is given by \cite{Blennow:2007pu}
\begin{equation}
P(\nu_\mu \to \nu_\mu;L) \simeq 1 - \sin^2(2\tilde{\theta}_{23}) \sin^2\left( \frac{\Delta \tilde{m}_{31}^2}{4E} L \right) \,,
\label{eq:Pmutau}
\end{equation}
where three-flavor effects due to $\Delta m_{21}^2$ and $\theta_{13}$ have been neglected, and the effective para\-meters are
\begin{eqnarray}
\Delta \tilde{m}_{31}^2 &=& \Delta m_{31}^2 \xi \,, \label{eq:deltatildem}\\
\sin^2(2\tilde{\theta}_{23}) &=& \frac{\sin^2(2\theta_{23})}{\xi^2} \label{eq:sintildetheta}
\end{eqnarray}
with
\begin{equation}
\xi = \sqrt{\left[ \hat{A} \varepsilon_{\tau\tau} + \cos(2\theta_{23}) \right]^2 + \sin^2(2\theta_{23})} \,. \label{eq:xi}
\end{equation}
Note that in equations~(\ref{eq:deltatildem})--(\ref{eq:xi}) we have used the NSI parameter $\varepsilon_{e\tau}$ as a perturbation and the formulae should hold if $|\varepsilon_{e\tau}|^2 A^2 L^2/(4 E^2) \ll 1$ or, in the case of the MINOS experiment, if $|\varepsilon_{e\tau}| \ll 5.8$ \cite{Blennow:2007pu}. In addition, note that equations~(\ref{eq:Pmutau})--(\ref{eq:xi}) are two-flavor neutrino oscillation formulae, which have been derived assuming $\Delta m_{21}^2 = 0$ and $\theta_{13} = 0$. Therefore, using equations~(\ref{eq:approxm})--(\ref{eq:approxV3}), which are three-flavor approximate mappings for small parameters $\alpha$, $s_{13}$ and $\varepsilon_{\alpha\beta}$, it is not directly possible to use them to derive equations~(\ref{eq:deltatildem})--(\ref{eq:xi}). Instead the results will be three-flavor approximations corresponding to the two-flavor formulae given in equations~(\ref{eq:deltatildem})--(\ref{eq:xi}). Furthermore, it is possible to show the effective three-flavor mixing matrix element $\tilde{U}_{e3}$ is given by
\begin{equation}
\tilde{U}_{e3} \simeq U_{e3} + \hat{A} \varepsilon_{e\tau} \cos (\theta_{23}) \,,
\end{equation}
which means that there could be a degeneracy between the mixing angle $\theta_{13}$ and the NSI parameter $\varepsilon_{e\tau}$ \cite{Blennow:2007pu}. Now, since the mixing angle $\theta_{13}$ has been measured \cite{An:2012eh,Abe:2011fz,Abe:2012tg,Ahn:2012nd}, the MINOS experiment can put a limit on $|\varepsilon_{e\tau}|$ \cite{Blennow:2007pu}. In fact, using data from the MINOS and T2K experiments, the bound $|\varepsilon_{e\tau}| \leq 1.3$ at 90\% C.L.~has been set \cite{Coelho:2012bp}. In addition to the above discussion for the MINOS experiment, it has recently been argued in the literature that it should be possible to study the NSI parameter $\varepsilon_{\mu\tau}$ using the MINOS experiment too \cite{Mann:2010jz,Kopp:2010qt,Isvan:2011fa}. In this case (assuming $\theta_{23} = 45^\circ$), the $\nu_\mu$ survival probability becomes \cite{Mann:2010jz}
\begin{equation}
P(\nu_\mu \to \nu_\mu;L) \simeq 1 - \sin^2 \left( \left| \frac{\Delta m_{31}^2}{4E} - \varepsilon_{\mu\tau} \frac{A}{2E} \right| L \right) \,. \label{eq:Pmmmt}
\end{equation}
Note that the amplitude of the second term is equal to 1, since $\theta_{23} = 45^\circ$ (maximal mixing) has been assumed, see \cite{Mann:2010jz} for details. Now, using a model based on the $\nu_\mu$ survival probability in equation~(\ref{eq:Pmmmt}) together with data from the MINOS experiment, the MINOS collaboration has obtained the following result for the matter NSI parameter $\varepsilon_{\mu\tau}$ at 90\% C.L. \cite{Coelho:2012}
$$
-0.200 < \varepsilon_{\mu\tau} < 0.070 \,,
$$
which means that MINOS has found no evidence for matter NSIs in its neutrino data, at least not a non-zero value for the matter NSI parameter $\varepsilon_{\mu\tau}$.

Second, in the case of the OPERA experiment, the important NSI parameter is $\varepsilon_{\mu\tau}$ (see equation~(\ref{eq:3prop})) due to the relatively short baseline length and the interesting transition probability is the appearance probability for oscillations of $\nu_\mu$ into $\nu_\tau$, which is given by \cite{Blennow:2008ym}
\begin{equation}
P(\nu_\mu \to \nu_\tau;L) = \left| c_{13}^2 \sin(2\theta_{23}) \frac{\Delta m_{31}^2}{4E} + \varepsilon_{\mu\tau}^* \frac{A}{2E} \right|^2 L^2 + {\cal O}(L^3) \,,
\end{equation}
where it has been assumed that the small mass-squared difference $\Delta m_{21}^2 = 0$. Thus, there exists a degeneracy between the fundamental neutrino oscillation parameters and the NSI parameter $\varepsilon_{\mu\tau}$. Note that it has been shown that the OPERA experiment is not very sensitive to the NSI parameters $\varepsilon_{e\tau}$ and $\varepsilon_{\tau\tau}$ \cite{EstebanPretel:2008qi}.

\subsection{Reactor neutrino experiments}

To my knowledge, NSIs in reactor neutrino experiments have only been discussed in \cite{Kopp:2007ne,Ohlsson:2008gx,Leitner:2011aa}. Below, we will summarize these three works.

First, in \cite{Kopp:2007ne}, a combined study on the performance of reactor and superbeam neutrino experiments in the presence of NSIs is presented. Indeed, in this work, the authors argue that reactor and superbeam data can be used to establish the presence of NSIs.

\begin{figure}
\includegraphics[width=\textwidth,bb = 25 530 550 740]{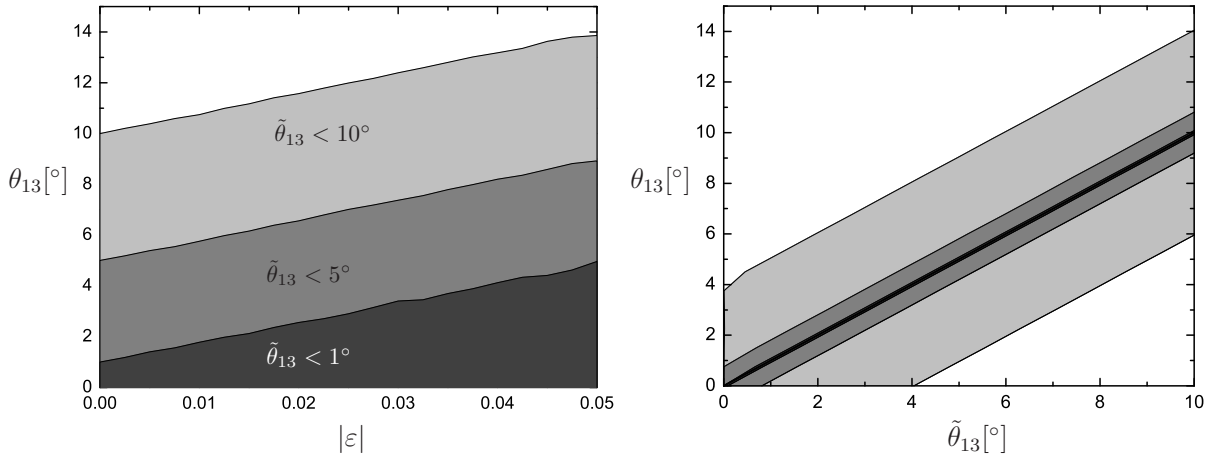}
\vspace{-5mm}
\caption{Mappings among $\tilde\theta_{13}$, $\theta_{13}$ and NSI parameters $\varepsilon_{\alpha\beta}$. In the left plot, the gray-shaded areas correspond to the indicated upper bounds on $\tilde\theta_{13}$, whereas in the right plot, the gray-shaded areas represent $|\varepsilon| < 0.05$ (light gray), $|\varepsilon| < 0.01$ (middle gray), and $|\varepsilon| < 0.001$ (dark gray), respectively. This figure has been reproduced with permission from \protect\cite{Ohlsson:2008gx}.}
\label{fig:theta13mappings}
\end{figure}

Second, in \cite{Ohlsson:2008gx}, NSIs at reactor neutrino experiments only were studied. In figure~\ref{fig:theta13mappings}, mappings among the effective mixing angle $\tilde\theta_{13}$, the fundamental mixing angle $\theta_{13}$ and the NSI parameters $\varepsilon_{\alpha\beta}$ are plotted. Without loss of generality, it is assumed that $|\varepsilon| \equiv |\varepsilon_{e\mu}| = |\varepsilon_{e\tau}|$.
Such plots could be important for the analyses of, for example, the Daya Bay, Double Chooz and RENO experiments. It was found that (i) $\theta_{13} < 14^\circ$, which is larger than the former CHOOZ bound that is about $10^\circ$ and also larger than the recently measured values of the mixing angle $\theta_{13}$ and (ii) inspite of a very small $\theta_{13}$, a sizable effective mixing angle can be achieved due to mimicking effects \cite{Ohlsson:2008gx}. In principle, this means that the measured value for the mixing angle $\theta_{13} \approx 9^\circ$ by the Daya Bay, Double Chooz and RENO experiments \cite{An:2012eh,Abe:2011fz,Abe:2012tg,Ahn:2012nd} could be a combination of the fundamental value for $\theta_{13}$ (which should be smaller than the effective measured value) and effects of NSI parameters.

\begin{figure}
  \includegraphics[width=0.5\textwidth]{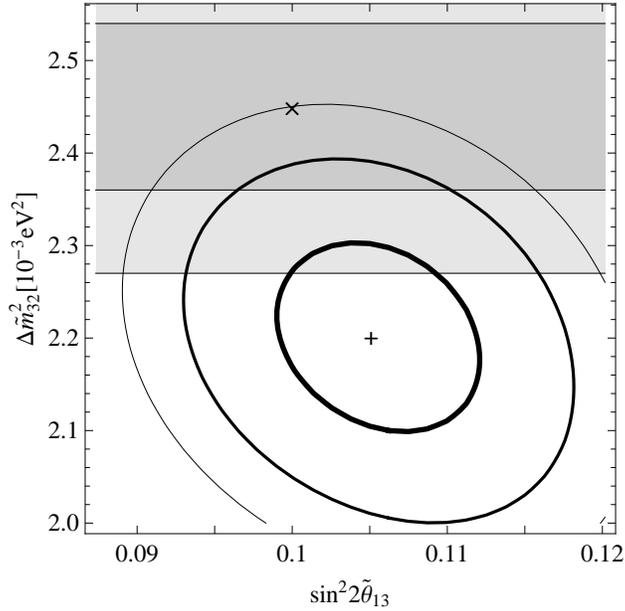}
  \vspace{-2.5mm}
  \caption{Effects of NSIs on $\theta_{13}$ and $\Delta m_{32}^2$ for the Daya Bay experiment after three years of running. The symbol `$\times$' denotes the assumed `true' values of the standard neutrino oscillations parameters, whereas the symbol `$+$' denotes the situation with production and detection NSIs included ($|\varepsilon| = 0.02$). The three curves show the $\chi^2$ levels around the best-fit point `$+$', respectively, where $\chi^2 = 20$ (thick, inner curve), $\chi^2 = 40$ (middle curve) and $\chi^2 = 60$ (thin, outer curve). The gray-shaded areas depict the `pull' by the large mass-squared difference $\Delta \tilde{m}_{32}^2$ departing from its `true' value, where the dark/light boundary encloses $\Delta \tilde{m}_{32}^2 = (2.45 \pm 0.09) \times 10^{-3} \, {\rm eV}^2$ and the light/white boundary encloses $\Delta \tilde{m}_{32}^2 = (2.45 \pm 0.18) \times 10^{-3} \, {\rm eV}^2$. This figure has been reproduced with permission from \cite{Leitner:2011aa}.}
  \label{fig:DayaBayNSIs}
\end{figure}

Third, in \cite{Leitner:2011aa}, NSIs at the Daya Bay experiment were studied. The authors of this work show that, under certain conditions, only three years of running of the Daya Bay experiment might be sufficient to provide a hint on production and detection NSIs. Thus, in future analyses of data from the Daya Bay experiment, it will be important to disentangle effects of NSI parameters on the mixing angle $\theta_{13}$ as well as the large neutrino mass-squared difference $\Delta m_{31}^2$. In figure~\ref{fig:DayaBayNSIs}, the effects of NSIs on $\theta_{13}$ and $\Delta m_{32}^2 \simeq \Delta m_{31}^2$ are shown for Daya Bay after three years of running.
We observe that production and detection NSIs give rise to a larger value of the effective leptonic mixing angle $\tilde\theta_{13}$ and a smaller value of the effective large mass-square difference $\Delta \tilde{m}_{32}^2$. For example, turning on the NSI parameters $|\varepsilon| \equiv |\varepsilon^{\rm s}_{e\alpha}| = |\varepsilon^{\rm d}_{\alpha e}| = 0.02$ ($\alpha = \mu,\tau$), the value of $\sin^2(2\tilde\theta_{13})$ is shifted from 0.1 to 0.105, whereas the value of $\Delta \tilde{m}_{32}^2$ is changed from $2.45 \cdot 10^{-3} \, {\rm eV}^2$ to $2.2 \cdot 10^{-3} \, {\rm eV}^2$.

\subsection{Neutrino Factory}
\label{sub:nf}

Due to the large sensitivity at a future neutrino factory for small parameters such as NSI parameters, there exist a vast amount of investigations for different neutrino factory setups in connection to NSIs \cite{Huber:2001zw,Huber:2001de,Huber:2002bi,Davidson:2003ha,Blennow:2005qj,Kopp:2007mi,Ribeiro:2007ud,Kopp:2008ds,Antusch:2008tz,Malinsky:2008qn,Tang:2009na,Gago:2009ij,Ohlsson:2009vk,Meloni:2009cg,Coloma:2011rq}. In this section, we will present most of these investigations and what their general conclusions for NSIs at a future neutrino factory are. Several investigations have discussed the sensitivity and discovery reach of a neutrino factory in probing NSIs \cite{Huber:2001zw,Huber:2001de,Huber:2002bi,Kopp:2007mi,Ribeiro:2007ud,Coloma:2011rq}. For example, it has been suggested that (i) a 100~GeV neutrino factory could probe flavor-changing neutrino interactions of the order of $|\varepsilon| \lesssim 10^{-4}$ at 99\% C.L. \cite{Huber:2001zw}, (ii) there is an entanglement between the mixing angle $\theta_{13}$ and NSI parameters\footnote{This could be excluded due to the non-zero and quite large value for the mixing angle $\theta_{13}$ found by the Daya Bay, Double Chooz and RENO collaborations  \cite{An:2012eh,Abe:2011fz,Abe:2012tg,Ahn:2012nd}.}, which would be solved in the best way by using the appearance channel $\nu_e \to \nu_\mu$ \cite{Huber:2001de,Huber:2002bi}, (iii) there are degeneracies between CP violation and NSI parameters \cite{Gago:2009ij,Coloma:2011rq}, (iv) a neutrino factory has excellent prospects in detecting NSIs originating from new physics at the TeV scale \cite{Kopp:2007mi} and (v) off-diagonal NSI parameters could be tested down to the order of $10^{-3}$, whereas diagonal NSI parameter combinations such as $\varepsilon_{ee} - \varepsilon_{\tau\tau}$ and $\varepsilon_{\mu\mu} - \varepsilon_{\tau\tau}$ could only be tested down to $10^{-1}$ and $10^{-2}$, respectively \cite{Coloma:2011rq}. Furthermore, there are studies on the optimization of a neutrino factory with respect to NSI parameters, especially for $\varepsilon_{\mu\tau}$ and $\varepsilon_{\tau\tau}$ \cite{Kopp:2008ds}, as well as the impact of near detectors (with $\nu_\tau$ detection) at a neutrino factory on NSIs \cite{Tang:2009na}. In addition, it has been pointed out that the generation of matter NSIs might give rise to production and detection NSIs at a neutrino factory \cite{Antusch:2008tz}. On the more theoretical side, some investigations of NSIs stemming from different models have been carried out. For example, NSIs from a triplet seesaw model \cite{Malinsky:2008qn} or the Zee--Babu model \cite{Ohlsson:2009vk} have been investigated. At a neutrino factory, NSIs from the triplet seesaw model could lead to quite significant signals of lepton flavor violating decays, whereas production NSIs from the Zee--Babu model might be at an observable level in the $\nu_e \to \nu_\tau$ and/or $\nu_\mu \to \nu_\tau$ channels. Finally, it has been pointed out that NSIs and non-unitarity might phenomenologically lead to very similar effects for a neutrino factory, although for completely different reasons \cite{Meloni:2009cg}.

\subsection{NSI effects on solar and supernova neutrino oscillations}
\label{sub:NSI_sol_sup}

In addition to studies on NSIs with atmospheric and man-made sources of neutrinos, there exist some investigations of NSI effects on solar and supernova neutrino oscillations \cite{Bergmann:2000gp,Berezhiani:2001rt,Guzzo:2004ue,Bolanos:2008km,Palazzo:2009rb,Das:2010sd,Palazzo:2011vg,Garces:2011aa,Agarwalla:2012wf,EstebanPretel:2007yu,Blennow:2008er,EstebanPretel:2009is,Dasgupta:2010ae,Das:2011gb}. First, we focus on NSIs with solar neutrinos, and then, we discuss NSIs with supernova neutrinos.

In the case of solar neutrinos, an analysis of data from the Super-Kamiokande and SNO experiments was performed to investigate the sensitivity of NSIs \cite{Berezhiani:2001rt}. Furthermore, it has been suggested that the Borexino experiment can provide signatures for NSIs \cite{Berezhiani:2001rt}, and its data can also be used to place constraints on NSIs \cite{Agarwalla:2012wf}. The LENA proposal has also been illustrated as a probe for NSIs \cite{Garces:2011aa}. In addition, non-universal flavor-conserving couplings with electrons, flavor-changing interactions and NSIs in general can affect the phenomeno\-logy of solar neutrinos \cite{Bergmann:2000gp,Guzzo:2004ue,Bolanos:2008km,Palazzo:2009rb,Das:2010sd,Palazzo:2011vg}.

In the case of supernova neutrinos, a three-flavor analysis for the possibility of probing NSIs, using neutrinos from a galactic supernova that propagate in the supernova envelope, has been performed \cite{EstebanPretel:2007yu}. Furthermore, the interplay between collective effects and NSIs for supernova neutrinos has been investigated \cite{Blennow:2008er,EstebanPretel:2009is,Dasgupta:2010ae}. Finally, a study on NSIs similar to \cite{Das:2010sd} for solar neutrinos has been presented for supernova neutrinos by the same authors~\cite{Das:2011gb}.

In addition to NSIs with solar and supernova neutrinos, other studies on NSIs with astrophysics have been carried out. For example, in \cite{Blennow:2009rp}, the authors investigate production and detection NSIs of high-energy neutrinos at neutrino telescopes, using neutrino flux ratios.

\section{Phenomenological bounds on NSIs}
\label{sec:NSIexp}

As discussed above, there are basically two neutrino experiments that have put direct bounds on NSI parameters---the Super-Kamiokande and MINOS experiments (see sections~\ref{sub:atmosne} and \ref{sub:accne}). However, there exist also some phenomenological works that have used different sets of data to find direct bounds on NSI parameters. Below, we will present direct bounds on both (i) matter NSIs and (ii) production and detection NSIs from such works. In addition, we will discuss bounds on NSIs in neutrino cross-sections as well as bounds on NSIs using accelerators.

\subsection{Direct bounds on matter NSIs}
\label{sub:dbmNSIs}

In a work by Davidson {\it et al} \cite{Davidson:2003ha}, bounds on matter NSI parameters using experiments with neutrinos and charged leptons (which are the LSND \cite{Auerbach:2001wg}, CHARM \cite{Dorenbosch:1986tb}, CHARM II \cite{Vilain:1994qy} and NuTeV \cite{Zeller:2001hh} experiments as well as data from LEP II \cite{Berezhiani:2001rs}) have been derived for the realistic scenario with three flavors \cite{Davidson:2003ha,Abdallah:2003np,Ribeiro:2007ud}
$$
\left( \begin{matrix} -0.9 < \varepsilon_{ee} < 0.75 & |\varepsilon_{e\mu}| \lesssim 3.8 \cdot 10^{-4} & |\varepsilon_{e\tau}| \lesssim 0.25 \\ & -0.05 < \varepsilon_{\mu\mu} < 0.08 & |\varepsilon_{\mu\tau}| \lesssim 0.25 \\ & & |\varepsilon_{\tau\tau}| \lesssim 0.4 \end{matrix} \right) \,.
$$
We observe that the bounds range from $10^{-4}$ to $1$ for the different matter NSI parameters. Note that the bounds presented in this analysis are obtained using loop effects. However, it turns out that bounds coming from loop effects (i.e.~one-loop level contributions to the four-charged-lepton interactions at tree level) are generally not applicable, since such bounds will be model dependent \cite{Biggio:2009kv}. Thus, in order to obtain non-ambiguous bounds, a gauge-invariant realization of the NSIs must be used. Therefore, in \cite{Biggio:2009kv}, Biggio {\it et al} have performed a new analysis. The result of this analysis is that the model-independent bound for the NSI parameter $\varepsilon_{e\mu}$ increases by a factor of $10^3$. It should be noted that one-loop effects on NSIs have also been studied in \cite{Bellazzini:2010gn}. Now, in \cite{Biggio:2009nt}, using bounds on NSI parameters from \cite{Davidson:2003ha,Barranco:2005ps,Barranco:2007ej,Bolanos:2008km}, but disregarding the loop bound on the parameter $\varepsilon_{e\mu}^{fC}$, bounds on the effective matter NSI parameters have been estimated by Biggio, Blennow and Fern{\'a}ndez-Mart{\'i}nez. Approximately, the bounds on the $\varepsilon_{\alpha\beta}$'s given in equation~(\ref{eq:matterNSIs}) are found to be
\begin{eqnarray}
\varepsilon_{\alpha\beta}^\oplus &\simeq& \sqrt{\sum_C \left[ \left( \varepsilon_{\alpha\beta}^{eC} \right)^2 + \left( 3 \varepsilon_{\alpha\beta}^{uC} \right)^2 + \left( 3 \varepsilon_{\alpha\beta}^{dC} \right)^2 \right]} \,, \\
\varepsilon_{\alpha\beta}^\odot &\simeq& \sqrt{\sum_C \left[ \left( \varepsilon_{\alpha\beta}^{eC} \right)^2 + \left( 2 \varepsilon_{\alpha\beta}^{uC} \right)^2 + \left( \varepsilon_{\alpha\beta}^{dC} \right)^2 \right]} \,,
\end{eqnarray}
where $\varepsilon_{\alpha\beta}^\oplus$ are the bounds for neutral Earth-like matter (with an equal number of proton and neutrons) and $\varepsilon_{\alpha\beta}^\odot$ are the bounds for neutral solar-like matter (consisting mostly of protons and electrons). Thus, the model-independent bounds on the matter NSI parameters are given by
\begin{eqnarray}
&& \left( \begin{matrix} |\varepsilon_{ee}| < 4.2 & |\varepsilon_{e\mu}| < 0.33 & |\varepsilon_{e\tau}| < 3.0 \\ & |\varepsilon_{\mu\mu}| < 0.068 & |\varepsilon_{\mu\tau}| < 0.33 \\ & & |\varepsilon_{\tau\tau}| < 21\end{matrix} \right) \quad \mbox{(Earth)} \,, \nonumber\\
&& \left( \begin{matrix} |\varepsilon_{ee}| < 2.5 & |\varepsilon_{e\mu}| < 0.21 & |\varepsilon_{e\tau}| < 1.7 \\ & |\varepsilon_{\mu\mu}| < 0.046 & |\varepsilon_{\mu\tau}| < 0.21 \\ & & |\varepsilon_{\tau\tau}| < 9.0\end{matrix} \right) \quad \mbox{(solar)} \,, \nonumber
\end{eqnarray}
which, except for the matter NSI parameters $\varepsilon_{\mu\mu}$ and $\varepsilon_{\mu\tau}$ in neutral solar-like matter, are larger than the too stringent bounds found by Davidson {\it et al} \cite{Davidson:2003ha}, and ranging between $10^{-2}$ and $10$, i.e.~they are one or two orders of magnitude larger than the previous bounds.

\subsection{Direct bounds on production and detection NSIs}

Finally, in \cite{Biggio:2009nt}, the authors have also derived model-independent bounds on production and detection NSIs. The most stringent bounds for charged-current-like NSI parameters for terrestrial experiments are the following
\begin{eqnarray}
&& \left( \begin{matrix} |\varepsilon_{ee}^{\mu e}| < 0.025 & |\varepsilon_{e\mu}^{\mu e}| < 0.030 & |\varepsilon_{e\tau}^{\mu e}| < 0.030 \\ |\varepsilon_{\mu e}^{\mu e}| < 0.025 & |\varepsilon_{\mu\mu}^{\mu e}| < 0.030 & |\varepsilon_{\mu\tau}^{\mu e}| < 0.030 \\ |\varepsilon_{\tau e}^{\mu e}| < 0.025 & |\varepsilon_{\tau \mu}^{\mu e}| < 0.030 & |\varepsilon_{\tau\tau}^{\mu e}| < 0.030 \end{matrix} \right) \,, \nonumber\\ && \nonumber\\
&& \left( \begin{matrix} |\varepsilon_{ee}^{ud}| < 0.041 & |\varepsilon_{e\mu}^{ud}| < 0.025 & |\varepsilon_{e\tau}^{ud}| < 0.041 \\ |\varepsilon_{\mu e}^{ud}| < \left\{\begin{array}{c} 1.8 \cdot 10^{-6} \\ 0.026 \end{array}\right. & |\varepsilon_{\mu\mu}^{ud}| < 0.078 & |\varepsilon_{\mu\tau}^{ud}| < 0.013 \\ |\varepsilon_{\tau e}^{ud}| < \left\{\begin{array}{c} 0.087 \\ 0.12 \end{array}\right. & |\varepsilon_{\tau \mu}^{ud}| < \left\{\begin{array}{c} 0.013 \\ 0.018 \end{array}\right. & |\varepsilon_{\tau\tau}^{ud}| < 0.13 \end{matrix} \right) \,, \nonumber
\end{eqnarray}
where, whenever two values are presented, the upper value refers to left-handed NSI parameters (i.e.~$\varepsilon_{\alpha\beta}^{udL}$) and the lower one to right-handed NSI parameters (i.e.~$\varepsilon_{\alpha\beta}^{udR}$), otherwise the value refers to both left- and right-handed NSI parameters (i.e.~$\varepsilon_{\alpha\beta}^{\mu eC}$ or $\varepsilon_{\alpha\beta}^{udC}$, where $C = L,R$). All these bounds are basically of the order of $10^{-2}$.

\subsection{Bounds on NSIs in neutrino cross-sections}
\label{sub:NSI_xsec}

In this section, NSIs in neutrino cross-sections will be investigated in detail. Neutrino NSIs with either electrons or first generation quarks can be constrained by low-energy scattering data. In general, one finds that bounds are stringent for muon neutrino interactions, loose for electron neutrino interactions, and in principle, do not exist for tau neutrino interactions. Note that in the present overview of the upper bounds on the NSI parameters, the results from Biggio, Blennow and Fern{\'a}ndez-Mart{\'i}nez \cite{Biggio:2009nt} have not been included.

\begin{figure}
  \includegraphics[height=.4\textheight]{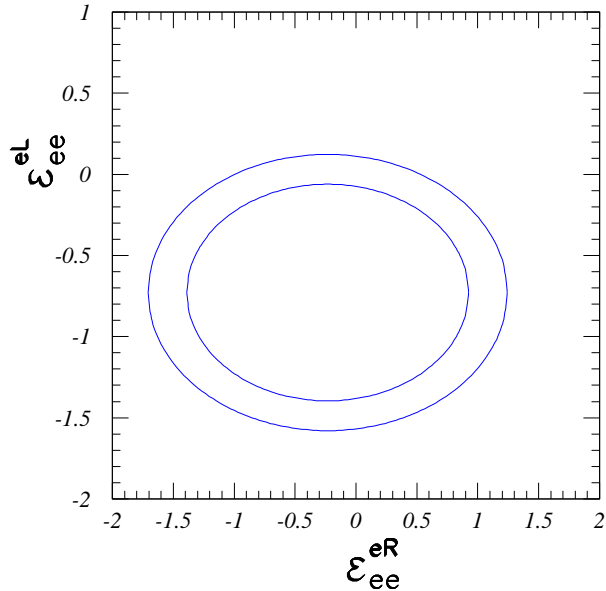}
  \vspace{-5mm}
  \caption{Bounds on flavor-conserving NSIs of $\nu_e e$ scattering from the LSND experiment. The area between the two ellipses is the allowed 90\%~C.L.~region. This figure has been reproduced with permission from \cite{Davidson:2003ha}.}
  \label{fig:LSND}
\end{figure}

The best measurement on electron neutrino--electron scattering comes from the LSND experiment that found the cross-section of this process to be \cite{Auerbach:2001wg}
\begin{eqnarray}
\sigma(\nu_e e \to \nu e) &=& \left[1.17 \pm 0.13 \, \mbox{(stat.)} \pm 0.12 \, \mbox{(syst.)} \right] \frac{G_F^2 m_e E_\nu}{\pi} \nonumber\\
&=& (1.17 \pm 0.17) \frac{G_F^2 m_e E_\nu}{\pi} \,,
\end{eqnarray}
where $m_e$ is the electron mass and $E_\nu$ is the neutrino energy, which should be compared with the SM cross-section \cite{Auerbach:2001wg}
\begin{eqnarray}
\left.\sigma(\nu_e e \to \nu e)\right|_{\rm SM} &=& \frac{2 G_F^2 m_e E_\nu}{\pi} \left[ \left(1+g_L^e\right)^2 + \frac{1}{3} \left(g_R^e\right)^2 \right] \nonumber\\
&=& \frac{1}{2} \left( 1 + 4 \sin^2 \theta_W + \frac{16}{3} \sin^4 \theta_W \right) \frac{G_F^2 m_e E_\nu}{\pi} \simeq 1.1048 \frac{G_F^2 m_e E_\nu}{\pi} \,, \label{eq:SMxsec}
\end{eqnarray}
where we have used the SM couplings of Z bosons and electrons $g_L^e = -\tfrac{1}{2} + \sin^2 \theta_W \simeq - 0.2688$ and $g_R^e = \sin^2 \theta_W \simeq 0.2312$ with $\sin^2 \theta_W = 0.23116 \pm 0.00012$ being the weak-mixing angle (or the Weinberg angle) \cite{Beringer:1900zz}. However, including electroweak radiative corrections\footnote{For an outline how the radiative corrections for electron neutrino--electron scattering can be computed, see appendix A in \cite{Bahcall:1995mm}.}, we obtain $g_L^e \simeq -0.2718$ and $g_R^e \simeq 0.2326$, which means that equation~(\ref{eq:SMxsec}) changes to
$\left.\sigma(\nu_e e \to \nu e)\right|_{\rm SM} \simeq 1.0967 G_F^2 m_e E_\nu/\pi$ \cite{Davidson:2003ha}.
Including NSIs, the expression for this cross-section becomes \cite{Davidson:2003ha,Forero:2011zz}
\begin{equation}
\sigma(\nu_e e \to \nu e) = \frac{2 G_F^2 m_e E_\nu}{\pi} \left[ (1 + g_L^e + \varepsilon_{ee}^{eL})^2 + \sum_{\alpha \neq e} |\varepsilon_{\alpha e}^{eL}|^2 + \frac{1}{3} (g_R^e + \varepsilon_{ee}^{eR})^2 + \frac{1}{3} \sum_{\alpha \neq e} |\varepsilon_{\alpha e}^{eR}|^2 \right] \,.
\end{equation}
Using the LSND data and the NSI cross-section for electron neutrino--electron scattering, we obtain 90\%~C.L.~bounds on the NSI parameters (note that only one NSI parameter at a time is considered) \cite{Davidson:2003ha}
$$
-0.07 < \varepsilon_{ee}^{eL} < 0.11 \,, \quad
-1 < \varepsilon_{ee}^{eR} < 0.5
$$
for flavor-conserving diagonal NSI parameters and
$$
|\varepsilon_{\tau e}^{eL}| < 0.4 \,, \quad |\varepsilon_{\tau e}^{eR}| < 0.7
$$
for flavor-changing NSI parameters. Considering both left- and right-handed diagonal NSIs, i.e.~two NSI parameters simultaneously, a 90\%~C.L.~region between two ellipses is obtained
$$
0.445 < (0.7282 + \varepsilon_{ee}^{eL})^2 + \frac{1}{3} (0.2326 + \varepsilon_{ee}^{eR})^2 < 0.725 \,,
$$
which is shown in figure~\ref{fig:LSND}.
In an updated phenomenological analysis \cite{Forero:2011zz} of data on electron neutrino--electron scattering including NSIs, more restrictive allowed 90\%~C.L.~bounds on the NSI parameters $\varepsilon_{ee}^{eL}$ and $\varepsilon_{ee}^{eR}$ have been found (considering both parameters simultaneously)
$$
-0.02 < \varepsilon_{ee}^{eL} < 0.09 \,, \quad
-0.11 < \varepsilon_{ee}^{eR} < 0.05 \,.
$$
However, note that the bounds on the two NSI parameters $\varepsilon_{ee}^{eL}$ and $\varepsilon_{ee}^{eR}$ in \cite{Davidson:2003ha} were derived considering only one NSI parameter at a time, whereas the corresponding bounds in \cite{Forero:2011zz} were found considering both parameters simultaneously. Therefore, the two sets of bounds are not directly comparable. In fact, it might be more reasonable to compare the bounds in \cite{Forero:2011zz} with the region between the two ellipses presented in figure~\ref{fig:LSND}.

Then, we turn our attention to electron neutrino--quark scattering. The CHARM collaboration \cite{Dorenbosch:1986tb} has measured the ratio between cross-sections and found
\begin{equation}
R^e = \frac{\sigma(\nu_e N \to \nu X) + \sigma(\bar\nu_e N \to \bar\nu X)}{\sigma(\nu_e N \to e X) + \sigma(\bar\nu_e N \to \bar e X)} = (\tilde g_L^e)^2 + (\tilde g_R^e)^2 = 0.406 \pm 0.140 \,.
\end{equation}
Including NSIs, the quantities $\tilde g_L^e$ and $\tilde g_R^e$ can be expressed as
\begin{align}
(\tilde g_L^e)^2 &= (g_L^u + \varepsilon_{ee}^{uL})^2 + \sum_{\alpha \neq e} |\varepsilon_{\alpha e}^{uL}|^2 + (g_L^d + \varepsilon_{ee}^{dL})^2 + \sum_{\alpha \neq e} |\varepsilon_{\alpha e}^{dL}|^2 \,, \\
(\tilde g_R^e)^2 &= (g_R^u + \varepsilon_{ee}^{uR})^2 + \sum_{\alpha \neq e} |\varepsilon_{\alpha e}^{uR}|^2 + (g_R^d + \varepsilon_{ee}^{dR})^2 + \sum_{\alpha \neq e} |\varepsilon_{\alpha e}^{dR}|^2 \,.
\end{align}
Using the CHARM data, we obtain 90\%~C.L.~bounds on the NSI parameters (only one NSI parameter at a time) \cite{Davidson:2003ha}
$$
-1 < \varepsilon_{ee}^{uL} < 0.3 \,, \quad
-0.3 < \varepsilon_{ee}^{dL} < 0.3 \,, \quad
-0.4 < \varepsilon_{ee}^{uR} < 0.7 \,, \quad
-0.6 < \varepsilon_{ee}^{dR} < 0.5
$$
for flavor-conserving NSIs and 
$$
|\varepsilon_{\tau e}^{qC}| < 0.5 \,, \quad q = u,d \quad \mbox{and} \quad C = L,R
$$
for flavor-changing NSIs. Again, considering all four NSI parameters simultaneously, a 90\%~C.L.~region is obtained
$$
0.176 < (0.3493 + \varepsilon_{ee}^{uL})^2 + (-0.4269 + \varepsilon_{ee}^{dL})^2 + (-0.1551 + \varepsilon_{ee}^{uR})^2 + (0.0775 + \varepsilon_{ee}^{dR})^2 < 0.636 \,,
$$
which describes two four-dimensional ellipsoids. In this case, note that the allowed 90\%~C.L.~region for the NSI parameters $\varepsilon_{ee}^{uL}$, $\varepsilon_{ee}^{dL}$, $\varepsilon_{ee}^{uR}$ and $\varepsilon_{ee}^{dR}$ is the four-dimensional space between the two ellipsoids.

\begin{figure}
  \includegraphics[height=.4\textheight]{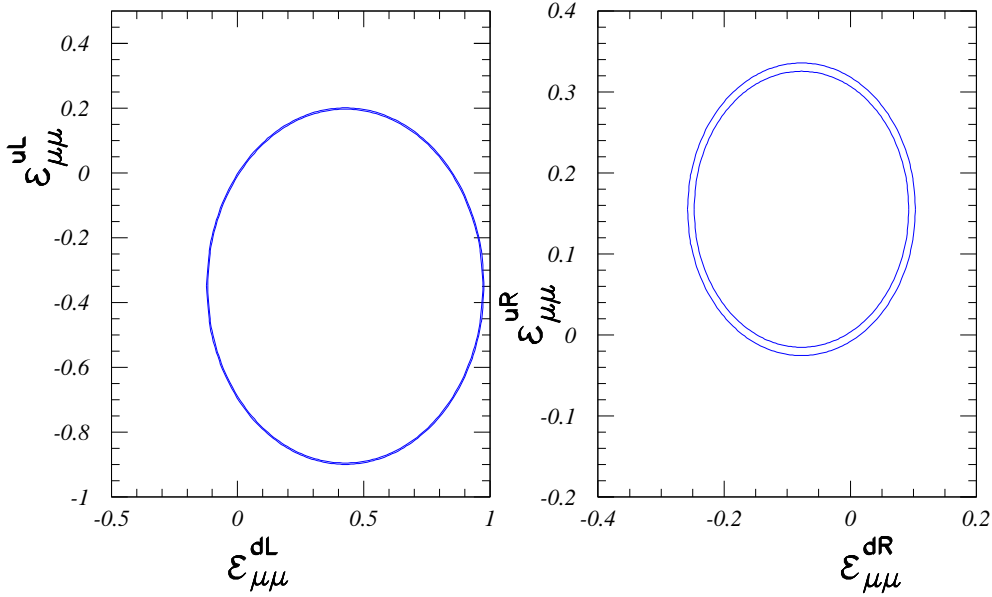}
  \vspace{-5mm}
  \caption{Bounds on flavor-conserving NSIs of $\nu_\mu q$ scattering from the NuTeV experiment. This figure has been reproduced with permission from \cite{Davidson:2003ha}.}
  \label{fig:nutev}
\end{figure}

Next, for muon neutrino--electron scattering, the CHARM II collaboration \cite{Vilain:1994qy} has measured $g_V^e = -0.035 \pm 0.017$ and $g_A^e = -0.503 \pm 0.017$, which translate into $g_L^e = -0.269 \pm 0.017$ and $g_R^e = 0.234 \pm 0.017$. Using these values, one can compute 90\%~C.L.~bounds on the NSI parameters (only one NSI parameter at a time) \cite{Davidson:2003ha}
$$
-0.025 < \varepsilon_{\mu\mu}^{eL} < 0.03 \,, \quad
-0.027 < \varepsilon_{\mu\mu}^{eR} < 0.03
$$
for flavor diagonal NSIs and
$$
|\varepsilon_{\tau\mu}^{eC}| < 0.1 \,, \quad C = L,R \nonumber
$$
for flavor-changing NSIs. Furthermore, the NuTeV collaboration \cite{Zeller:2001hh} has measured $(\tilde g_L^\mu)^2 = 0.3005 \pm 0.0014$ and $(\tilde g_R^\mu)^2 = 0.0310 \pm 0.0011$ that appear in ratios of cross-sections for neutrino--nucleon (muon neutrino--quark) scattering processes. Using these values, we obtain (only one NSI parameter at a time) \cite{Davidson:2003ha}
\begin{eqnarray}
&& -0.009 < \varepsilon_{\mu\mu}^{uL} < -0.003 \,, \quad
0.002 < \varepsilon_{\mu\mu}^{dL} < 0.008 \,, \nonumber\\
&& -0.008 < \varepsilon_{\mu\mu}^{uR} < 0.003 \,, \quad
-0.008 < \varepsilon_{\mu\mu}^{dR} < 0.015 \nonumber
\end{eqnarray}
for flavor diagonal NSIs and
$$
|\varepsilon_{\tau\mu}^{qR}| < 0.05 \,, \quad q = u,d
$$
for flavor-changing NSIs. Similarly, the 90\%~C.L.~regions using two
NSI parameters simultaneously are presented in figure~\ref{fig:nutev}.

Finally, we investigate NSIs for the $e^+ e^- \to \nu \bar\nu \gamma$ cross-section at LEP II. The 90\%~C.L.~bounds on flavor diagonal NSIs are given by \cite{Berezhiani:2001rs}
$$
-0.6 < \varepsilon_{\tau\tau}^{eL} < 0.4 \,, \quad
-0.4 < \varepsilon_{\tau\tau}^{eR} < 0.6 \,,
$$
whereas for flavor-changing NSIs
$$
| \varepsilon_{\alpha\beta}^{eC}| < 0.4 \,, \quad C = L,R, \quad \alpha = \tau, \quad \beta = e,\mu \,.
$$
Also, for the NSI parameters $\varepsilon_{\tau\tau}^{eL}$ and $\varepsilon_{\tau\tau}^{eR}$, there exists an update \cite{Forero:2011zz} of the 90\%~C.L.~bounds
$$
-0.51 < \varepsilon_{\tau\tau}^{eL} < 0.34 \,, \quad
-0.35 < \varepsilon_{\tau\tau}^{eR} < 0.50 \,.
$$

In conclusion, using neutrino cross-section measurements for the determination of upper bounds on NSIs, the only non-zero parameters obtained are $\varepsilon_{\mu\mu}^{uL}$ and $\varepsilon_{\mu\mu}^{dL}$. Therefore, for future analyses, new data on neutrino cross-sections would be valuable. In fact, there exist several experiments that will provide improved measurements on neutrino cross-sections in the future. For example, the ArgoNeuT \cite{Spitz:2010dj}, MINER$\nu$A \cite{McFarland:2006}, MiniBooNE/SciBooNE \cite{Cheng:2012yy}, MINOS \cite{Adamson:2009ju} and T2K \cite{Abe:2011} experiments have already delivered data. In addition, there is an interesting proposed future experiment called nuSTORM \cite{Kyberd:2012iz}, which could perform precision neutrino cross-section measurements that will be important for future long-baseline neutrino oscillation experiments.

\subsection{Bounds on NSIs using accelerators}
\label{sub:acc}

An interesting alternative option to neutrino experiments is to use data from accelerators to study NSIs, and especially, bounds on NSIs. Examples of accelerators that can be used are the LEP collider (1989--2000), the Tevatron (1987--2011) and the LHC (2009--present). Basically, this option would mean a potential interplay between neutrino experiments and accelerators, and the topical importance of the LHC should, of course, be used in connection to neutrino physics.

First, in \cite{Davidson:2011kr,Davidson:2011xz}, the authors have explored the alternative option to study NSIs by bringing into play collider data. Indeed, searching for so-called dimension-6 contact interactions in the channels $e^+ e^- \to e^+ e^-, \mu^+ \mu^-, \tau^+ \tau^-, \bar{q} q$ (where $q = u,d,s,c,b$), the data from LEP II provides bounds on these interactions \cite{Schael:2006wu,Abbiendi:2003dh}, which in turn can be used to set bounds on NSIs of the order of $\varepsilon \lesssim (10^{-2} - 10^{-3})$ at $\sqrt{s} \sim 200$~GeV. Furthermore, if NSIs are contact interactions at LHC energies,  such contact interactions would induce $q \bar{q} \to W^+ W^- \ell_\alpha^+ \ell_\beta^-$ and the LHC should have a sensitivity reach for NSIs of the order of $\varepsilon \gtrsim 3 \times 10^{-3}$ at $\sqrt{s} = 14$~TeV and with 100~${\rm fb}^{-1}$ of data. Second, assuming the NSIs remain contact at LHC energies and using monojet plus missing transverse-energy data for the processes $q \bar{q} \to \overline{\nu_\alpha} \nu_\beta j$ (where $j = g,q,\bar{q}$) to probe NSIs, bounds on the NSI parameters $\varepsilon_{ee}^{qC}$, $\varepsilon_{\tau\tau}^{qC}$ and $\varepsilon_{\tau e}^{qC}$ ($q = u,d$ and $C = L,R$) have been derived \cite{Friedland:2011za}, which are of the order of $\varepsilon \lesssim (0.1 - 1)$ at $\sqrt{s} = 7$~TeV, using $1~{\rm fb}^{-1}$ data from the ATLAS experiment \cite{Aad:2011xw} at the LHC. Note that the ATLAS data are already superior to the data from the CDF experiment at the Tevatron.
In addition, using $4.98~{\rm fb}^{-1}$ data from the CMS experiment \cite{Chatrchyan:2012mea} at the LHC, bounds on NSI parameters induced by $q \bar{q} \to W^+ W^- \ell_\alpha^+ \ell_\beta^-$ are also found to be of the order of $\varepsilon \lesssim (0.1 - 1)$ at $\sqrt{s} = 7$~TeV \cite{Friedland:2011za}.

With the advent of the new data from the LHC (2009--2012), it would be interesting to search for signals of NSIs in these data. So far, no evidence for physics beyond the SM have been observed at the LHC. Therefore, I urge the ATLAS and CMS collaborations at the LHC as well as phenomenologists to investigate NSIs as signals for new physics. Especially, using the results of the phenomenological analyses \cite{Davidson:2011kr,Davidson:2011xz,Friedland:2011za}, the processes (i) $pp \to j \overline{\nu_\alpha} \nu_\beta$ (where $j = g,q,\bar{q}$) and (ii) $p p \to W^+ W^- \ell_\alpha^+ \ell_\beta^-$ look promising to search for NSIs at the LHC.

\section{Outlook for NSIs}
\label{sec:NSIdiscovery}

I am sure that more investigations on NSIs will be conducted in the future, both theoretical and experimental ones. In addition, I believe that a future neutrino factory\footnote{Note that an interesting and important question is if a neutrino factory is going to be built or not, given a non-zero and quite large value for the mixing angle $\theta_{13}$.} would be the most plausible experiment to find signatures of NSIs (see the detailed discussion in section~\ref{sub:nf}), but perhaps also the LHC will shed some light (see section~\ref{sub:acc}). The power of a neutrino factory is that it may have a sensitivity and discovery reach for the NSI parameters. Concerning existing and running neutrino oscillation experiments, the discussed experiments in section~\ref{sec:NSIpheno} probably belong to the ones that could obtain hints on NSIs. To repeat, such experiments are the Super-Kamiokande, MINOS, T2K and the reactor neutrino experiments. However, it is rather unlikely that they will be sensitive to NSIs, and they certainly lack any discovery reach. Considering future conventional neutrino experiments, a near detector (that can measure taus) placed close to ordinary man-made neutrino sources would be a feasible setup to observe production and detection NSIs. Future long-baseline experiments should possibly be designed to be able to probe NSIs. Finally, an interesting new option is to perform neutrino oscillation experiments with the ESS that is planned to be built in Sweden, and such experiments might have sensitivities to NSIs, but need to be investigated in detail. In order to summarize this outlook, the most promising experiments to search for NSIs are (i) a future neutrino factory, (ii) the LHC and (iii) various long-baseline experiments.

\section{Summary and conclusions}
\label{sec:S&C}

In summary of this review, we have discussed NSIs as sub-leading effects to the standard paradigm for neutrino flavor transitions based on the phenomenon of neutrino oscillations. In particular, we have presented both (i) production and detection NSIs including the so-called zero-distance effect as well as (ii) matter NSIs. In the case of matter NSIs, we have given approximate analytical model-independent mappings between effective neutrino masses and leptonic mixing angles and the fundamental neutrino mass-squared differences and leptonic mixing parameters. In addition, we have studied approximate two-flavor formulae for neutrino flavor transitions. Furthermore, we have presented some different theoretical models for NSIs such as a seesaw model and the Zee--Babu model, and investigated the phenomenology of NSIs in general. In particular, we have displayed the experimental results of upper bounds on NSIs from the analyses of the Super-Kamiokande ($|\varepsilon_{\mu\tau}| < 0.033, |\varepsilon_{\tau\tau} - \varepsilon_{\mu\mu}| < 0.147$) and MINOS ($-0.200 < \varepsilon_{\mu\tau} < 0.070$) collaborations, which both mean that no evidence for matter NSIs has been found. Moreover, we have indicated the sensitivities of NSIs for accelerators, a future neutrino factory and the reactor neutrino experiment Daya Bay. For example, mimicking effects induced by NSIs could play a very important role in reactor neutrino experiments, especially for the mixing angle $\theta_{13}$. In fact, the fundamental value for $\theta_{13}$ could be smaller than the measured value due to an interplay with NSIs. Finally, we have presented phenomenological upper bounds on NSIs. In the case of matter NSIs, former results gave bounds ranging from $10^{-4}$ to $1$, whereas latter results yielded bounds between $10^{-2}$ and $10$. In the case of production and detection NSIs for terrestrial experiments, there are bounds of the order from $10^{-6}$ to $0.1$. In addition, we have given bounds for NSIs using data from neutrino cross-section measurements. Indeed, low-energy neutrino scattering experiments measuring neutrino cross-sections can be used to set bounds on NSI parameters. In conclusion, we have shortly discussed an outlook for the future sensitivity and discovery reach of NSIs, which could be responsible for neutrino flavor transitions on a sub-leading level. Especially, the LHC, a future neutrino factory and long-baseline experiments could open a new window towards determining the possible NSIs.

\begin{acknowledgments}
The author would like to thank Mattias Blennow and He Zhang for useful discussions and comments. This work was supported by the Swedish Research Council (Vetenskapsr{\aa}det), contract no. 621-2011-3985 (T.O.).
\end{acknowledgments}


\appendix
\section{Abbreviations}

\begin{table}[!h]
{\scriptsize
\begin{tabular}{ll}
\hline
{\em General abbreviations}\\
CERN & European Organization for Nuclear Research, Geneva, Switzerland (see section~\ref{sub:accne})\\
C.L. & confidence level\\
CP & charge parity\\
ESS & European Spallation Source, Lund, Sweden (see section~\ref{sec:NSIdiscovery})\\
Fermilab & Fermi National Accelerator Laboratory, Batavia, Illinois, USA (see section~\ref{sub:accne})\\
GUT & Grand Unified Theory\\
LEP & Large Electron-Positron collider @ CERN (see section~\ref{sub:acc})\\
LHC & Large Hadron Collider @ CERN (see section~\ref{sec:intro})\\
LNGS & Laboratori Nazionali del Gran Sasso, Gran Sasso, Italy (see section~\ref{sub:acc})\\
MSW & Mikheyev--Smirnov--Wolfenstein\\
NSI & non-standard neutrino interaction\\
SM & Standard Model\\
{\em Experiments} \\
ATLAS & A Toroidal LHC ApparatuS (see section~\ref{sub:acc})\\
Borexino & The name Borexino is an Italian diminutive of BOREX (Boron solar neutrino experiment)\\
 & (see section~\ref{sub:NSI_sol_sup})\\
CHARM & CERN-Hamburg-Amsterdam-Rome-Moscow (see section~\ref{sub:NSI_xsec})\\
ICARUS & Imaging Cosmic And Rare Underground Signals (see section~\ref{sec:intro})\\
K2K & KEK (High Energy Accelerator Research Organization) to Kamioka (see section~\ref{sec:intro})\\
KamLAND & Kamioka Liquid scintillator AntiNeutrino Detector (see section~\ref{sec:intro})\\
KATRIN & KArlsruhe TRItium Neutrino (see section~\ref{sec:intro})\\
LENA & Low Energy Neutrino Astronomy (see section~\ref{sub:NSI_sol_sup})\\
LSND & Liquid Scintillator Neutrino Detector (see section~\ref{sub:NSI_xsec})\\
MiniBooNE & Mini Booster Neutrino Experiment (see section~\ref{sec:intro})\\
MINOS & Main Injector Neutrino Oscillation Search (see section~\ref{sec:intro})\\
NO$\nu$A & NuMI Off-Axis $\nu_e$ Appearance (see section~\ref{sec:intro})\\
NuTeV & Neutrinos at the Tevatron (see section~\ref{sub:NSI_xsec})\\
OPERA & Oscillations Project with Emulsion-tRacking Apparatus (see section~\ref{sec:intro})\\
RENO & Reactor Experiment for Neutrino Oscillations (see section~\ref{sec:intro})\\
SNO & Sudbury Neutrino Observatory (see section~\ref{sec:intro})\\
Super-Kamiokande & Super-Kamioka Neutrino Detection Experiment (see section~\ref{sec:intro})\\
T2K & Tokai to Kamioka (see section~\ref{sec:intro})\\
\hline
\end{tabular}
}
\end{table}

\clearpage


%

\end{document}